\documentclass[12pt]{article}
\usepackage[dvips]{graphicx}
\usepackage{float}
\usepackage{latexsym}
\usepackage{amssymb}
\usepackage{amsbsy}
\usepackage{amsmath}   
\usepackage{amsfonts}

\renewcommand{\theequation}{\arabic{section}.\arabic{equation}}

\makeatletter
\@addtoreset{equation}{section}
\makeatother

\def\p{\partial}
\def\dfrac#1#2{{\displaystyle\frac{#1}{#2}}}
\def\stTD#1#2{\hbox to 0em{\mathsurround=0em $\stackrel{#1}{\makebox[0pt]{} #2}$\hss} \phantom{#2}}\def\stscript#1#2{\hbox to 0em{\mathsurround=0em ${\scriptstyle\stackrel{#1}{\makebox[0pt]{} #2}}$\hss} \phantom{#2}}\def\stscriptscript#1#2{\hbox to 0em{\mathsurround=0em ${\scriptscriptstyle\stackrel{#1}{\makebox[0pt]{} #2}}$\hss} \phantom{#2}}
\def\comb#1#2#3{{\mathsurround 0pt\hbox to 0pt {\hspace*{#3}\raisebox{#2}{${#1}$}\hss}}}
\def\combs#1#2#3{{\mathsurround 0pt\hbox to 0pt {\hspace*{#3}\raisebox{#2}{${\scriptstyle #1}$}\hss}}}
\def\combss#1#2#3{{\mathsurround 0pt\hbox to 0pt {\hspace*{#3}\raisebox{#2}{${\scriptscriptstyle #1}$}\hss}}}
\def\e#1{\mathrm{e}^{#1}}

\def\df{\mathrm{d}}

\def\metrEff{\mathchoice{\combs{\sim}{1ex}{0.2ex}\mathfrak{m}}{\combs{\sim}{1ex}{0.2ex}\mathfrak{m}}{\combss{\sim}{0.66ex}{0.05ex}\mathfrak{m}}{}{}}
\def\metr{\mathfrak{m}}
\def\bcdot{\mathchoice{\mathbin{\boldsymbol{\comb{\cdot}{0.22ex}{0.57ex}\Diamond}}}{\mathbin{\boldsymbol{\comb{\cdot}{0.22ex}{0.57ex}\Diamond}}}{\mathbin{\boldsymbol{\combs{\cdot}{0.13ex}{0.35ex}\Diamond}}}{}{}}
\def\bwedge{\mathchoice{\mathbin{\combs{\boldsymbol{\backslash}}{0.2ex}{0.1ex}\boldsymbol{\wedge}}}{\mathbin{\combs{\boldsymbol{\backslash}}{0.2ex}{0.1ex}\boldsymbol{\wedge}}}{\mathbin{\combss{\boldsymbol{\backslash}}{0.1ex}{0ex}\boldsymbol{\wedge}}}{}{}}
\def\Eem{E}
\def\bEem{\mathbf{E}}
\def\Hem{H}
\def\bHem{\mathbf{H}}
\def\Dem{D}
\def\bDem{\mathbf{D}}
\def\Bem{B}
\def\bBem{\mathbf{B}}
\def\baab{\mathbf{b}}
\def\bbaab{{\mathsurround 0pt\mbox{${\bf b}$\hspace*{-1.1ex}${\bf b}$}}}
\def\p{\partial}

\def\unitc{\mathbf{1}}
\def\him{\imath}
\def\bhim{\boldsymbol{\imath}}
\def\hconj#1{\mathchoice{{{}^{\boldsymbol{*}}\mspace{-3.7mu}#1}}{{{}^{\boldsymbol{*}}\mspace{-3.5mu}#1}}{{{}^{\boldsymbol{*}}\mspace{-4mu}#1}}{}{}}
\def\bRe{\boldsymbol{\Re}}
\def\bIm{\boldsymbol{\Im}}
\def\bDB{\mathbf{Y}}
\def\bEH{\mathbf{Z}}

\def\bnp{\comb{\boldsymbol{\cdot}}{0ex}{0.4ex} {\boldsymbol{\partial}}}
\def\FourierPN#1#2#3{{}_{\mathtt{#2 f}}^{\;\;#3}\!\mspace{-1.5mu}#1}
\def\FourierF#1#2#3{{}_{\mathtt{#2}}^{\,#3}\!\mspace{-1.5mu}#1}

\def\cylaf#1#2{\mathchoice{\combs{\|}{0.42ex}{0.63ex}{\mathbf{C}^{#2}_{#1}}}{\combs{\|}{0.42ex}{0.63ex}{\mathbf{C}^{#2}_{#1}}}{\combss{\|}{0.245ex}{0.41ex}{\mathbf{C}^{#2}_{#1}}}{}{}}

\def\sphersecf#1#2{\mathchoice{\comb{\circ}{0.24ex}{0.13ex}{\mathbf{S}_{#1}^{#2}}}{\comb{\circ}{0.24ex}{0.13ex}{\mathbf{S}_{#1}^{#2}}}{\combs{\circ}{0.16ex}{0.12ex}{\mathbf{S}_{#1}^{#2}}}{}{}}
\def\spherzonf#1#2#3{\mathchoice{\comb{\circ}{0.24ex}{0.23ex}{\mathbf{Z}_{#1#3}^{#2}}}{\comb{\circ}{0.24ex}{0.24ex}{\mathbf{Z}_{#1#3}^{#2}}}{\combs{\circ}{0.15ex}{0.2ex}{\mathbf{Z}_{#1#3}^{#2}}}{}{}}
\def\spheraf#1#2#3{\mathchoice{\comb{\circ}{0.23ex}{0.5ex}{\mathbf{C}^{#2#3}_{#1}}}{\comb{\circ}{0.23ex}{0.5ex}{\mathbf{C}^{#2#3}_{#1}}}{\combs{\circ}{0.17ex}{0.38ex}{\mathbf{C}^{#2#3}_{#1}}}{}{}}
\def\RadfunS#1#2{\mathchoice{\combs{\boldsymbol{\uparrow}}{0.4ex}{0.2ex}{\mathbf{S}^{#1}_{#2}}}{\combs{\boldsymbol{\uparrow}}{0.4ex}{0.2ex}{\mathbf{S}^{#1}_{#2}}}{\combss{\boldsymbol{\uparrow}}{0.2ex}{0.1ex}{\mathbf{S}^{#1}_{#2}}}{}{}}

\def\garmsb#1#2#3#4{\mathchoice{{}_#2\!\combs{\comb{\circ}{-0.2ex}{0.06ex}{\approx}}{0.4ex}{0.41ex}{\mathbf{C}^{#3#4}_{#1}}}{{}_#2\!\combs{\comb{\circ}{-0.2ex}{0.06ex}{\approx}}{0.4ex}{0.41ex}{\mathbf{C}^{#3#4}_{#1}}}{{}_#2\!\combss{\combs{\circ}{-0.05ex}{0.13ex}{\approx}}{0.2ex}{0.25ex}{\mathbf{C}^{#3#4}_{#1}}}{}{}}
\def\cEp{\mathchoice{\combs{-}{0.3ex}{-0.1ex}\mathcal{E}}{\combs{-}{0.3ex}{-0.1ex}\mathcal{E}}{\combss{-}{0.25ex}{-0.12ex}\mathcal{E}}{}{}}
\def\cEt{\mathchoice{\combs{\sim}{0.3ex}{-0.1ex}\mathcal{E}}{\combs{\sim}{0.3ex}{-0.1ex}\mathcal{E}}{\combss{\sim}{0.25ex}{-0.12ex}\mathcal{E}}{}{}}
\def\cP{\mathcal{P}}
\def\bcP{\boldsymbol{\mathcal{P}}}
\def\cE{\mathcal{E}}
\def\bsoll#1{\mathchoice{\combs{\circ}{1.6ex}{0.7ex}{#1}}{\combs{\circ}{1.6ex}{0.7ex}{#1}}{\combss{\circ}{1.1ex}{0.45ex}{#1}}{}{}}
\def\lsolp#1#2{\mathchoice{\combs{\sim#2}{1.6ex}{-0.05ex}{#1}}{\combs{\sim#2}{1.6ex}{-0.05ex}{#1}}{\combss{\sim#2}{1.1ex}{-0.2ex}{#1}}{}{}}
\def\lsol#1{\mathchoice{\combs{\sim}{1.6ex}{0.45ex}{#1}}{\combs{\sim}{1.6ex}{0.45ex}{#1}}{\combss{\sim}{1.1ex}{0.22ex}{#1}}{}{}}
\def\Cn{\bar{\C}}
\def\Cm{\mathchoice{\combs{m}{0.45ex}{0.07em}\mathrm{C}}{\combs{m}{0.45ex}{0.07em}\mathrm{C}}{\combss{m}{0.25ex}{0.04em}\mathrm{C}}{}{}}
\def\Ce{\mathchoice{\combs{e}{0.45ex}{0.22em}\mathrm{C}}{\combs{e}{0.45ex}{0.22em}\mathrm{C}}{\combss{e}{0.3ex}{0.15em}\mathrm{C}}{}{}}
\def\C{\mathchoice{\comb{\cdot}{0.22ex}{0.26em}\mathrm{C}}{\comb{\cdot}{0.22ex}{0.26em}\mathrm{C}}{\combs{\cdot}{0.14ex}{0.18em}\mathrm{C}}{}{}}
\def\tomega{\tilde{\omega}}
\def\rt{\tilde{r}}
\def\rb{\bar{r}}
\def\Czero{\mathchoice{\combs{0}{0.25ex}{0.55ex}{C}}{\combs{0}{0.25ex}{0.55ex}{C}}{\combss{0}{0.1ex}{0.3ex}{C}}{}{}}
\def\Cinf{\mathchoice{\combs{\infty}{0.45ex}{0ex}{C}}{\combs{\infty}{0.45ex}{0ex}{C}}{\combss{\infty}{0.33ex}{-0.05ex}{C}}{}{}}
\def\ypmp#1#2{\phi_{#1}^{#2}}
\def\ypma#1#2{a_{#1}^{#2}}
\def\ypmat#1#2{\tilde{a}_{#1}^{#2}}
\def\ypm#1#2{y_{#1}^{#2}}
\def\zpm#1{z_{#1}}
\def\mmod#1{|#1|}
\def\marg#1{\langle #1\rangle}
\def\OOO#1#2{\mathcal{O}^{#1}_{#2}}
\def\ooo#1#2{o^{#1}_{#2}}
\def\polcf#1{\mathchoice{\combs{\uparrow}{0.3ex}{0.5ex}{O_{\!#1}}}{\combs{\uparrow}{0.3ex}{0.5ex}{O_{\!#1}}}{\combss{\uparrow}{0.2ex}{0.33ex}{O_{\!#1}}}{}{}}
\def\Cron#1#2{\mathchoice{\combs{\backslash}{0.1ex}{0.1ex}\delta^{#1}_{#2}}{\combs{\backslash}{0.1ex}{0.1ex}\delta^{#1}_{#2}}{\combss{\backslash}{0ex}{0ex}\delta^{#1}_{#2}}{}{}}
\def\resfunpp{\mathchoice{\combs{\prime\prime}{1.6ex}{0.8ex}\resfun}{\combs{\prime\prime}{1.6ex}{0.8ex}\resfun}{\combss{\prime\prime}{1.05ex}{0.5ex}\resfun}{}{}}
\def\resfun{\mathchoice{\combs{\bumpeq}{0.4ex}{0.3ex}Q}{\combs{\bumpeq}{0.4ex}{0.3ex}Q}{\combss{\bumpeq}{0.2ex}{0.15ex}Q}{}{}}
\def\sphcoefpw#1{\mathchoice{\combs{\circleddash}{0.4ex}{0.25ex}{C_{#1}}}{\combs{\circleddash}{0.4ex}{0.25ex}{C_{#1}}}{\combss{\circleddash}{0.2ex}{0.1ex}{C_{#1}}}{}{}}
\def\radfppp#1#2#3{\zeta_{#1}^{#2,#3}}

\def\average#1#2#3#4{\mathchoice{\left\langle\mspace{-#4 mu}\left| #1 \right|\mspace{-#4 mu}\right\rangle^{#2}_{#3}}{\left\langle\mspace{-#4 mu}\left| #1 \right|\mspace{-#4 mu}\right\rangle^{#2}_{#3}}{\left\langle\mspace{-#4 mu}\left| #1 \right|\mspace{-#4 mu}\right\rangle^{#2}_{#3}}{}{}}
\def\Vol{\mathchoice{\combs{\square}{0.2ex}{0.2ex}\mathrm{V}}{\combs{\square}{0.2ex}{0.2ex}\mathrm{V}}{\combss{\square}{0.12ex}{0.095ex}\mathrm{V}}{}{}}
\def\dVol{\mathchoice{\mathrm{d}\hspace{-0.3ex}\Vol}{\mathrm{d}\hspace{-0.3ex}\Vol}{\mathrm{d}\hspace{-0.3ex}\Vol}{}{}}
\def\refC#1#2{R^{#1}_{#2}}
\def\resC#1#2{\tilde{R}^{#1}_{#2}}

\def\mtitle{Linear waves around static dyon solution of nonlinear electrodynamics}

\def\mabstract{Nonlinear electrodynamics model in hypercomplex form is considered.
Its linearization around a solution is obtained. The appropriate problem for linear waves
around static dyon solution (SDS) of Born-Infeld electrodynamics is investigated.
Two types of wave scattering on SDS are considered: dissipative (with momentum transmission from plane wave
to SDS) and non-dissipative (for SDS imbedded to an equilibrium wave background).
Resonance phenomenon in the problem is discovered and some
resonance frequencies are obtained by using a numerical method.
 The form of resonance wave modes are discussed.
The sum of a plane wave (as the elementary component of the wave background)
with one resonance mode is considered.
The appropriate energy density is investigated at infinity.
The averaged energy density is demonstrated to have the term proportional
to inverse radius. This fact allow to consider such field configurations
as the cause of gravitational interaction, taking into account
the effective Riemann space effect discovered in my previous works. A behavior of
the linearized solution at origin of coordinates and the problem beyond the linearization are discussed.}

\begin{document}

\title{\textbf{\mtitle}}
\author{{\bf A. A. Chernitskii}\\
\small A. Friedmann Laboratory for Theoretical Physics, St.-Petersburg\\
\small and\\
\small State University of Engineering and Economics\\
\small Marata str. 27, St.-Petersburg, Russia, 191002\\
\small \texttt{AAChernitskii@engec.ru}
}
\date{}

\maketitle

\begin{abstract}
\mabstract
\end{abstract}

\section{Introduction}
\label{introd}
Let us use the accordance between electromagnetic particles and space localized solutions of suitable
electrodynamic field model.
These model solutions will be called particle ones.

Consider static particle solution of some electrodynamics model.
Such a solution is the space-localized static field configuration satisfying model equations.
For the usual linear electrodynamics we have a static particle solution in the form of Coulomb potential
that corresponds to electrically charged particle.
There are appropriate solutions for nonlinear electrodynamics. But in the case of nonlinear model the
static solution will modify conditions of propagation for small additional electromagnetic waves.
That is the space with a static solution looks like heterogeneous and anisotropic optical medium
for these waves.
The space-localized heterogeneity of a special kind, as with a static particle solution, must give a scattering effect.
This scattering may have a set of resonance frequencies determined by a specific size of the
heterogeneity region.

Dyon particle solution has point singularity with electric and magnetic charges.
Here I consider a nonlinear model known as Born-Infeld electrodynamics and its dyon static solution
\cite{Chernitskii1999}.
This approach can help us to advance in solution of some fundamental physical problems
connected with particle physics.

A similar problem for scattering of light by electrically charged particle solution
of Born-Infeld electrodynamics was considered by
Schr{\"o}dinger \cite{Schrodinger1942a,Schrodinger1942b}, where the solution was seeking in the form
of power series in frequency of light. But this form of the solution can be appropriate only for small frequencies.
It should be noted that the corresponding mathematical problem is difficult for analytical investigation.

The present state of the art in computational modeling allows to succeed for
the problems, which still remained unsolved just because of its mathematical difficulty.

In the present work I use a method for investigation of the problem based on numerical calculations.
This method is suitable for arbitrary characteristics of the electromagnetic waves.

\section{Mathematical model\\for nonlinear electrodynamics}
\label{mathmod}
Nonlinear electrodynamics in hypercomplex or Clifford number form \cite{Chernitskii2002a} is the most convenient to use.
Here I consider the known representation for nonlinear electrodynamics, in which electric and magnetic inductions are used as
dependent variables (see also my paper \cite{Chernitskii1999}).
We have the following hypercomplex form for the basic equation without singularities \cite{Chernitskii2005a}:
\label{44396100}
\begin{align}
\label{dkmfjysg}
& \p_0\,\bDB + \bnp \bcdot \bDB + \bnp\bwedge\bEH = 0
\;\;,
\end{align}
where $\p_\mu\doteqdot {\p}/{\p x^\mu}$ (Greek indices take the values $0,1,2,3$) and
$\bnp\doteqdot\baab_0\,\baab^i\,\p_i$
(Latin indices take the values $1,2,3$) is the operator of space differentiation.
Here $\baab^{\mu}$ are hypercomplex numbers appropriate to basis vectors.
Symmetrical $\bcdot$ and asymmetrical $\bwedge$ products are constructed from noncommutative product for the hypercomplex numbers.
For coordinate systems with the components of metric $\metr_{00}=-1$ and $\metr_{0i} = 0$ we have $\bnp= \bbaab^{i}\,\p_i$, where
$\bbaab^{i}$ are basis bivectors. At the present paper I use a coordinate system of this type.

We have by definition
\begin{equation}
\label{YZ}
\bDB \doteqdot \bDem + \bhim\,\bBem
\;\;,\quad
\bEH \doteqdot \bEem + \bhim\,\bHem
\;\;,
\end{equation}
where $\bhim$ is hyperimaginary unit, $\bDem\doteqdot\Dem^{i}\,\bbaab_{i}$ and  $\bBem\doteqdot\Bem^{i}\,\bbaab_{i}$
are electric and magnetic inductions, $\bEem\doteqdot\Eem^{i}\,\bbaab_{i}$ and  $\bHem\doteqdot\Hem^{i}\,\bbaab_{i}$
are electric and magnetic intensities.
The quasi-bivectors of electromagnetic induction $\bDB$ and intensity $\bEH$ are invariant only for space transformations
of coordinate system but not for transformations affecting time.
The hyperimaginary unit $\bhim$ is permutable with $\p_0$, $\bDB$, $\bEH$, and $\bnp$.
Thus for this case it can be used as conventional imaginary unit.
The prefix ``hyper'' can be omitted in some words of the present paper.

Symmetrical product is connected with scalar product for space vectors, for example
\mbox{$\bDem\bcdot\bBem = \bDem\cdot\bBem$},
and asymmetrical one is connected with their vector product, for example
\mbox{$\bDem\bwedge\bBem = \bhim\,\bDem\times\bBem$}.
But symmetrical and asymmetrical products have more wide sense for hypercomplex numbers.
For the details see also my paper \cite{Chernitskii2005a}.

Equation (\ref{44396100}) must be supplemented by the constitutive relations, connecting
electromagnetic inductions with intensities: $\bDem=\bDem(\bEem,\bBem)$, $\bHem=\bHem(\bEem,\bBem)$.
This connection can be expressed with the help of one scalar function
$\cE=\cE (\bDem,\bBem)$, which is an energy density for the model.  We have for the components
\begin{equation}
\label{38187544}
\Eem_{i} = 4\pi\,\frac{\p \cE}{\p\Dem^{i}}
\quad,\qquad
\Hem_{i} = 4\pi\,\frac{\p \cE}{\p\Bem^{i}}
\;\;.
\end{equation}

In view of (\ref{YZ}) we can consider a function $\cE (\bDB,\hconj{\bDB})$ instead of $\cE (\bDem,\bBem)$.
Here $\hconj{\bDB}$ designates hyperconjugate hypernumber for $\bDB$, realized with substitution of
$\bhim$ by $-\bhim$ in canonical representation (\ref{YZ}).

We can consider strong or Frechet derivatives \cite{HutsonPym1980} of some function with respect
to $\bDB$ and $\hconj{\bDB}$,
where the bivector module $|\bDB|\doteqdot\sqrt{\bDem^2+\bBem^2}$ is considered as the appropriate norm.
These strong derivatives are linear operators working on a function placed on the right of the operator.
The strong derivative appears, in particular, in differentiation of function for vector argument.
For example,
\begin{equation}
\label{44468525}
\frac{\p}{\p\Dem^{i}}\,\cE (\bDB,\hconj{\bDB}) = \frac{\p \cE}{\p\bDB}:\frac{\p\bDB}{\p\Dem^{i}} +
\frac{\p \cE}{\p\hconj{\bDB}}:\frac{\p\hconj{\bDB}}{\p\Dem^{i}}
\;\;,
\end{equation}
where the symbol $:$ is introduced to distinct the operation from the noncommutative product.
If the strong derivative exists, then there is also a weak derivative equal to strong one.
The weak derivative is defined by the regular way:
\begin{equation}
\label{38182255}
\frac{\p \mathbf{f}(\bDB)}{\p\bDB}:\mathbf{Q} \doteqdot
\left[\frac{\p}{\p \varepsilon}\, \mathbf{f}(\bDB+\varepsilon\,\mathbf{Q})\right]_{\varepsilon=0}
\;\;.
\end{equation}

According to (\ref{38187544}), (\ref{YZ}), and (\ref{44468525}) we have
\begin{equation}
\label{52292250}
\frac{\p \cE}{\p \bDB}:\lsol{\bDB} = \frac{1}{8\pi}\,\hconj{\bEH}\bcdot\lsol{\bDB}
\quad,\qquad
\frac{\p \cE}{\p \hconj{\bDB}}:\hconj{\lsol{\bDB}} = \frac{1}{8\pi}\,\bEH\bcdot\hconj{\lsol{\bDB}}
\;\;.
\end{equation}

For Born-Infeld model (see also \cite{Chernitskii2004a}) we have the following expression for energy density
(see, for example, my paper\footnote{In the present paper the designations are somewhat changed.}
\cite{Chernitskii1999}):
\begin{subequations}
\begin{align}
\label{}
\cE &= \frac{\left(\cEt - 1\right)}{4\pi\,\chi^2}
\;\;,
\end{align}
where
\label{qlllssswwe}
\begin{align}
\label{77812269}
\cEt &\doteqdot\sqrt{1+8\pi\,\chi^2\,\cEp + (4\pi)^2\,\chi^4\,\bcP^2}
\;\;,\\
\label{alwqpk}
 \cEp &\doteqdot \frac{1}{8\pi}\,\bDB\bcdot\hconj{\bDB} = \frac{1}{8\pi}\left(\bDem^2 + \bBem^2\right)
\;\;,\\
\label{qlojjj}
\bcP &\doteqdot  \frac{1}{8\pi}\,\bDB\bwedge\hconj{\bDB} = -\frac{\bhim}{4\pi}\left(\bDem\bwedge \bBem\right)
= \frac{1}{4\pi}\left(\bDem\times \bBem\right)
\;\;,
\end{align}
$\chi$ is a dimensional constant of nonlinearity for the model.
Here $\cEp$ is energy density for linear electrodynamics, that is for $|\bDB|\to 0$.
$\bcP$ is momentum density both for linear and nonlinear electrodynamics.
It is obvious
\begin{equation}
\label{41855026}
\bcP\bcdot\bDB \equiv 0
\;\;.
\end{equation}
\end{subequations}

Having a form for the energy density $\cE (\bDB,\hconj{\bDB})$, we can write an appropriate nonlinear constitutive relation
according to (\ref{38187544}) and (\ref{YZ}):
\begin{equation}
\label{38314796}
\bEH=\bEH (\bDB,\hconj{\bDB})
\;\;.
\end{equation}
The constitutive relation specifies a concrete nonlinear electrodynamics model.

According to (\ref{qlllssswwe}), (\ref{38187544}), (\ref{YZ}) we have the following constitutive relation
for Born-Infeld model in hypercomplex form:
\begin{equation}
\label{63244197}
\bEH = \frac{\unitc + 4\pi\,\chi^2\,\bcP}{\cEt}\,\bDB
\;\;.
\end{equation}

This constitutive relation differs from one for the case where electric intensity $\bEem$ and magnetic induction
$\bBem$ are considered as dependent variables \cite{Chernitskii2002a}.

\section{Linearization around some solution}
\label{linnns}
Consider a solution $\bsoll{\bDB}$ for nonlinear electrodynamics model and a linearization $\lsol{\bDB}$ around $\bsoll{\bDB}$.
Thus I find a solution to equation (\ref{dkmfjysg}) in the form
\begin{equation}
\label{36671462}
\bDB = \bsoll{\bDB} + \lsol{\bDB}
\;\;,
\end{equation}
where $\lsol{\bDB}$ is small:
\begin{equation}
\label{45674616}
\mmod{\chi\,\lsol{\bDB}} \ll 1
\;\;.
\end{equation}
Here $\chi$ is considered as the dimensional constant for general electromagnetic field model with arbitrary
constitutive relation (\ref{38314796}).

Substitution of (\ref{36671462}) into constitutive relations (\ref{38314796}) and
consideration (\ref{45674616}) give the following linearization for the general case:
\begin{equation}
\label{37678841}
\bEH = \bsoll{\bEH} + \left.\frac{\p\bEH}{\p\bDB}\right|_{\bDB=\bsoll{\bDB}} : \lsol{\bDB}
 + \left.\frac{\p\bEH}{\p\hconj{\bDB}}\right|_{\bDB=\bsoll{\bDB}} : \hconj{\lsol{\bDB}}
\;\;,
\end{equation}
where $\bsoll{\bEH}\doteqdot\bEH(\bsoll{\bDB})$.

For relation (\ref{63244197}) we have
\begin{multline}
\label{39606993}
\bEH = \bsoll{\bEH} + \lsol{\bEH}
\quad,
\\
\lsol{\bEH} = \bsoll{\cEt}^{-1}\left\{\lsol{\bDB} +
(\chi^2/2)\left[\left(\bsoll{\bDB}\bwedge\hconj{\bsoll{\bDB}}\right)\bwedge\lsol{\bDB}+
\left(\bsoll{\bDB}\bwedge\hconj{\lsol{\bDB}}\right)\bwedge\bsoll{\bDB}+
\left(\lsol{\bDB}\bwedge\hconj{\bsoll{\bDB}}\right)\bwedge\bsoll{\bDB}
\right.\right.
\\
-\left.\left. \bsoll{\bEH}\left(\bsoll{\bEH}\bcdot\hconj{\lsol{\bDB}} + \hconj{\bsoll{\bEH}}\bcdot\lsol{\bDB}\right)\right]\right\}
\;\;.
\end{multline}
Here we can apply the following useful formula:
\begin{equation}
\label{42004190}
\left(\mathbf{B}\bwedge\mathbf{C}\right)\bwedge\mathbf{A} =
\mathbf{B}\left(\mathbf{A}\bcdot\mathbf{C}\right) - \mathbf{C}\left(\mathbf{A}\bcdot\mathbf{B}\right)
\;\;.
\end{equation}

\section{Static dyon solution (SDS)}
\label{sphstsol}
The simplest spherically symmetric static solution of Born-Infeld electrodynamics
has one singular point with electric and magnetic charges and called dyon \cite{Chernitskii1999}.
This static dyon solution (SDS) can be written in the following hypercomplex form:
\begin{equation}
\label{42909272}
\bsoll{\bDB} = \frac{\C}{r^2}\,\spheraf{0}{0}{0}
\quad,
\qquad
\bsoll{\bEH} = \frac{\C}{\sqrt{r^4 + \rb^{4}}}\,\spheraf{0}{0}{0}
\quad,
\end{equation}
where $\C\doteqdot \Ce + \bhim\,\Cm$ is electromagnetic charge, $\Ce$ is electric charge, $\Cm$ is magnetic one,
$\spheraf{j}{l}{m}$ are angle spherical bivector functions introduced in my paper \cite{Chernitskii2005a},
$\spheraf{0}{0}{0} = \bbaab_{r}$ is radial basis bivector, $\rb\doteqdot \sqrt{|\chi\,\C|}$ is a characteristic radius
of the static dyon solution.

Introduce also a normed charge $\Cn\doteqdot\dfrac{\C}{|\C|}$, which characterizes a ratio between
electric and magnetic charges.

\section{Linearization around SDS\\and appropriate solutions}
\label{lineqssol}
Let us consider a time harmonic field configuration expanded in terms of the angle spherical bivector functions
$\spheraf{j}{l}{m}$ \cite{Chernitskii2005a}:
\begin{equation}
\label{46364520}
\lsol{\bDB} = \Cn\,\sum_{j=-1}^{1}\,\sum_{l=|j|}^{\infty}\,\sum_{m=-l}^{l}\spheraf{j}{l}{m}\,
\left(\FourierPN{\lsol{\bDB}_{j}^{lm}}{+}\,\e{-\bhim\,\omega\,x^0} +
\FourierPN{\lsol{\bDB}_{j}^{lm}}{-}\,\e{\bhim\,\omega\,x^0} \right)
\;\;,
\end{equation}
where $\spheraf{j}{l}{m}=\spheraf{j}{l}{m} (\vartheta,\varphi)$,
$\FourierPN{\lsol{\bDB}_{j}^{lm}}{\pm}{}=\FourierPN{\lsol{\bDB}_{j}^{lm}}{\pm}{}(r)$,
$\omega>0$. Here the multiplication by the normed
charge $\Cn$ is used for a simplification of resulting formulas.
The analogous representation is considered for quasi-bivector $\lsol{\bEH}$.

\subsection{Equations}
\label{equations}
Substituting (\ref{46364520}) into (\ref{39606993}), taking into consideration (\ref{42909272}), and using the dependent variables
introduced in \cite{Chernitskii2005a}, I obtain
\begin{subequations}
\label{64537728}
\begin{align}
\label{64547649}
\FourierPN{\lsol{\bEH}_{+}^{lm}}{\pm}{} &= \frac{r^2}{\sqrt{r^4+\rb^4}}\left\{
\FourierPN{\lsol{\bDB}_{+}^{lm}}{\pm}{} + \frac{\rb^{4}}{2\,r^4}\left[\FourierPN{\lsol{\bDB}_{+}^{lm}}{\pm}{}+
(-1)^{m}\,\FourierPN{\lsol{\bDB}_{+}^{l,-m}}{\mp}{*}\right]
\right\}
\;\;,\\
\label{64549831}
\FourierPN{\lsol{\bEH}_{0}^{lm}}{\pm}{} &= \frac{r^2}{\sqrt{r^4+\rb^4}}\left\{
\FourierPN{\lsol{\bDB}_{0}^{lm}}{\pm}{} - \frac{\rb^{4}}{2\,(r^4 + \rb^4)}\left[\FourierPN{\lsol{\bDB}_{0}^{lm}}{\pm}{}+
(-1)^{m}\,\FourierPN{\lsol{\bDB}_{0}^{l,-m}}{\mp}{*}\right]\right\}
\;\;,\\
\label{64654396}
\FourierPN{\lsol{\bEH}_{-}^{lm}}{\pm}{} &= \frac{r^2}{\sqrt{r^4+\rb^4}}\left\{
\FourierPN{\lsol{\bDB}_{-}^{lm}}{\pm}{} + \frac{\rb^{4}}{2\,r^4}\left[\FourierPN{\lsol{\bDB}_{-}^{lm}}{\pm}{} -
(-1)^{m}\,\FourierPN{\lsol{\bDB}_{-}^{l,-m}}{\mp}{*}\right]\right\}
\;\;,
\end{align}
\end{subequations}
where
\vspace{-2ex}
\begin{equation}
\label{54051520}
\begin{array}{ccc}
&\FourierPN{\lsol{\bDB}_{+}^{lm}}{\pm}{} \doteqdot \FourierPN{\lsol{\bDB}_{1}^{lm}}{\pm}{} + \FourierPN{\lsol{\bDB}_{-1}^{lm}}{\pm}{}
\;\;,&
\quad\FourierPN{\lsol{\bDB}_{-}^{lm}}{\pm}{} \doteqdot \FourierPN{\lsol{\bDB}_{1}^{lm}}{\pm}{} - \FourierPN{\lsol{\bDB}_{-1}^{lm}}{\pm}{}
\;\;,
\\[1ex]
&\FourierPN{\lsol{\bEH}_{+}^{lm}}{\pm}{} \doteqdot \FourierPN{\lsol{\bEH}_{1}^{lm}}{\pm}{} + \FourierPN{\lsol{\bEH}_{-1}^{lm}}{\pm}{}
\;\;,&\quad
\FourierPN{\lsol{\bEH}_{-}^{lm}}{\pm}{} \doteqdot \FourierPN{\lsol{\bEH}_{1}^{lm}}{\pm}{} - \FourierPN{\lsol{\bEH}_{-1}^{lm}}{\pm}{}
\;\;.
\end{array}
\end{equation}

According to \cite{Chernitskii2005a} there is the following system of equations:
\begin{subequations}
\label{75771872}
\begin{align}
\label{75776575}
(r^2\,\FourierPN{\lsol{\bDB}_{0}^{lm}}{\pm}{})_{;r} + \bhim\,r\,\sqrt{l\,(l+1)}\, \FourierPN{\lsol{\bDB}_{+}^{lm}}{\pm}{} &= 0
\;\;,\\
\label{75779626}
(r\,\FourierPN{\lsol{\bEH}_{+}^{lm}}{\pm}{})_{;r} \mp \bhim\,r\,\omega\,\FourierPN{\lsol{\bDB}_{-}^{lm}}{\pm}{}
- \bhim\,\sqrt{l\,(l+1)}\, \FourierPN{\lsol{\bEH}_{0}^{lm}}{\pm}{} &= 0
\;\;,\\
\label{75783096}
(r\,\FourierPN{\lsol{\bEH}_{-}^{lm}}{\pm}{})_{;r} \mp \bhim\,r\,\omega\,\FourierPN{\lsol{\bDB}_{+}^{lm}}{\pm}{}  &= 0
\;\;,\\
\label{75785796}
r\,\omega\,\FourierPN{\lsol{\bDB}_{0}^{lm}}{\pm}{} \pm \sqrt{l\,(l+1)}\,\FourierPN{\lsol{\bEH}_{-}^{lm}}{\pm}{}   &= 0
\;\;.
\end{align}
\end{subequations}
where $(...)_{;r}$ is derivative with respect to radius $r$.

To simplify relations (\ref{64537728}) introduce again the following new variables:
\begin{equation}
\label{43850835}
\FourierPN{\lsol{\bDB}_{p}^{lm}}{\pm|}{} \doteqdot \FourierPN{\lsol{\bDB}_{p}^{lm}}{+}{} \pm (-1)^{m}\,\FourierPN{\lsol{\bDB}_{p}^{l,-m}}{-}{*}
\;,\qquad
\FourierPN{\lsol{\bEH}_{p}^{lm}}{\pm|}{} \doteqdot \FourierPN{\lsol{\bEH}_{p}^{lm}}{+}{} \pm (-1)^{m}\,\FourierPN{\lsol{\bEH}_{p}^{l,-m}}{-}{*}
\;,
\end{equation}
where $p=-,0,+$.

Inverting relations (\ref{54051520}) and (\ref{43850835}) we have
\begin{equation}
\label{37346145}
\begin{array}{rl}
\FourierPN{\lsol{\bDB}_{0}^{lm}}{+}{} &= \dfrac{1}{2}\,\Bigl(\FourierPN{\lsol{\bDB}_{0}^{lm}}{+|}{} + \FourierPN{\lsol{\bDB}_{0}^{lm}}{-|}{}\Bigr)
\;\;,
\\[1.5ex]
\FourierPN{\lsol{\bDB}_{0}^{lm}}{-}{} &=
\dfrac{(-1)^{m}}{2}\,\Bigl(\FourierPN{\lsol{\bDB}_{0}^{l,-m}}{+|}{*} - \FourierPN{\lsol{\bDB}_{0}^{l,-m}}{-|}{*}\Bigr)
\;\;,
\\[1.5ex]
\FourierPN{\lsol{\bDB}_{\pm 1}^{lm}}{+}{} &=
\dfrac{1}{4}\,\Bigl[\Bigl(\FourierPN{\lsol{\bDB}_{+}^{lm}}{+|}{} + \FourierPN{\lsol{\bDB}_{+}^{lm}}{-|}{}\Bigr)
\pm \Bigl(\FourierPN{\lsol{\bDB}_{-}^{lm}}{+|}{} + \FourierPN{\lsol{\bDB}_{-}^{lm}}{-|}{}\Bigr)\Bigr]
\;\;,
\\[1.5ex]
\FourierPN{\lsol{\bDB}_{\pm 1}^{lm}}{-}{} &=
\dfrac{(-1)^{m}}{4}\,\Bigl[\Bigl(\FourierPN{\lsol{\bDB}_{+}^{l,-m}}{+|}{*} - \FourierPN{\lsol{\bDB}_{+}^{l,-m}}{-|}{*}\Bigr)
\pm \Bigl(\FourierPN{\lsol{\bDB}_{-}^{l,-m}}{+|}{*} - \FourierPN{\lsol{\bDB}_{-}^{l,-m}}{-|}{*}\Bigr)\Bigr]
\;\;.
\end{array}\end{equation}

Using (\ref{43850835}) I obtain the following relations from (\ref{64537728}):
\begin{subequations}
\label{46557788}
\begin{align}
\label{46560772}
\FourierPN{\lsol{\bEH}_{+}^{lm}}{+|}{} &= \frac{\sqrt{r^4+\rb^4}}{r^2}\;\FourierPN{\lsol{\bDB}_{+}^{lm}}{+|}{}\;\;,
&
\FourierPN{\lsol{\bEH}_{+}^{lm}}{-|}{} &= \frac{r^2}{\sqrt{r^4+\rb^4}}\;\FourierPN{\lsol{\bDB}_{+}^{lm}}{-|}{}
\;\;,\\
\label{46564067}
\FourierPN{\lsol{\bEH}_{0}^{lm}}{+|}{} &= \frac{r^6}{(r^4+\rb^4)^{3/2}}\;\FourierPN{\lsol{\bDB}_{0}^{lm}}{+|}{}\;\;,
&
\FourierPN{\lsol{\bEH}_{0}^{lm}}{-|}{} &= \frac{r^2}{\sqrt{r^4+\rb^4}}\;\FourierPN{\lsol{\bDB}_{0}^{lm}}{-|}{}
\;\;,\\
\label{46566540}
\FourierPN{\lsol{\bEH}_{-}^{lm}}{+|}{} &= \frac{r^2}{\sqrt{r^4+\rb^4}}\;\FourierPN{\lsol{\bDB}_{-}^{lm}}{+|}{}\;\;,
&
\FourierPN{\lsol{\bEH}_{-}^{lm}}{-|}{} &= \frac{\sqrt{r^4+\rb^4}}{r^2}\;\FourierPN{\lsol{\bDB}_{-}^{lm}}{-|}{}
\end{align}
\end{subequations}
and the following equation system from (\ref{75771872}):
\begin{subequations}
\label{65903676}
\begin{align}
\label{65905929}
(r^2\,\FourierPN{\lsol{\bDB}_{0}^{lm}}{\pm|}{})_{;r} + \bhim\,r\,\sqrt{l\,(l+1)}\, \FourierPN{\lsol{\bDB}_{+}^{lm}}{\mp|}{} &= 0
\;\;,\\
\label{65908533}
(r\,\FourierPN{\lsol{\bEH}_{+}^{lm}}{\pm|}{})_{;r} - \bhim\,r\,\omega\,\FourierPN{\lsol{\bDB}_{-}^{lm}}{\pm|}{}
- \bhim\,\sqrt{l\,(l+1)}\, \FourierPN{\lsol{\bEH}_{0}^{lm}}{\mp|}{} &= 0
\;\;,\\
\label{65912299}
(r\,\FourierPN{\lsol{\bEH}_{-}^{lm}}{\pm|}{})_{;r} -\bhim\,r\,\omega\,\FourierPN{\lsol{\bDB}_{+}^{lm}}{\pm|}{}  &= 0
\;\;,\\
\label{65914842}
r\,\omega\,\FourierPN{\lsol{\bDB}_{0}^{lm}}{\pm|}{} + \sqrt{l\,(l+1)}\,\FourierPN{\lsol{\bEH}_{-}^{lm}}{\mp|}{}   &= 0
\;\;.
\end{align}
\end{subequations}

Differentiating (\ref{65905929}) and making necessary substitutions from (\ref{46557788}) and (\ref{65903676})
I obtain the following two equations for the functions $\FourierPN{\lsol{\bDB}_{0}^{lm}}{\pm|}{} (r)$:
\begin{subequations}
\label{42952668}
\begin{align}
\label{42971430}
\left\{\frac{\df^2}{\df r^2} + \frac{2\left(r^{4}+2\,\rb^{4}\right)}{r\left(r^{4}+\rb^{4}\right)}\,\frac{\df}{\df r}
+ \left[\omega^2 - \frac{l\,(l+1)\,r^{4}-2\,\rb^{4}}{r^2\left(r^{4}+\rb^{4}\right)}\right]\right\}
\left(r\,\FourierPN{\lsol{\bDB}_{0}^{lm}}{+|}{}\right) &= 0
\;,\\
\label{42973663}
\left\{\frac{\df^2}{\df r^2} + \frac{2\,r^{3}}{\left(r^{4}+\rb^{4}\right)}\,\frac{\df}{\df r}
+ \left[\omega^2 - \frac{l\,(l+1)\,r^{4}+2\,\rb^{4}}{r^2\left(r^{4}+\rb^{4}\right)}\right]\right\}
\left(r\,\FourierPN{\lsol{\bDB}_{0}^{lm}}{-|}{}\right) &= 0
\;.
\end{align}
\end{subequations}

These equations can be written also in the form
\begin{subequations}
\label{44531721}
\begin{align}
\label{44533615}
\left\{\frac{\df^2}{\df r^2} + \frac{2\,\rb^{4}}{r\left(r^{4}+\rb^{4}\right)}\,\frac{\df}{\df r}
+ \left[\omega^2 - \frac{l\,(l+1)\,r^{2}}{\left(r^{4}+\rb^{4}\right)}\right]\right\}
\left(r^2\,\FourierPN{\lsol{\bDB}_{0}^{lm}}{+|}{}\right) &= 0
\;,\\
\label{44536089}
\left\{\frac{\df^2}{\df r^2} - \frac{2\,\rb^{4}}{r\left(r^{4}+\rb^{4}\right)}\,\frac{\df}{\df r}
+ \left[\omega^2 - \frac{l\,(l+1)\,r^{2}}{\left(r^{4}+\rb^{4}\right)}\right]\right\}
\left(r^2\,\FourierPN{\lsol{\bDB}_{0}^{lm}}{-|}{}\right) &= 0
\;.
\end{align}
\end{subequations}

We can introduce the following dimensionless radius and frequency:
\begin{equation}
\label{49802800}
\rt \doteqdot \frac{r}{\rb}
\quad,\qquad
\tomega \doteqdot \omega\,\rb
\quad,\qquad
\tomega\,\rt \equiv \omega\,r
\quad.
\end{equation}
The appropriate ``dimensionless'' equations are obtained with the substitution $\rb\to 1$, $\omega\to\tomega$, $r\to\rt$
into equations (\ref{42952668}) and (\ref{44531721}).

The operators of equations (\ref{44531721}) coincide with operators of
equations considered by Schr{\"o}dinger \cite{Schrodinger1942a,Schrodinger1942b}.
But dependent variables in Schr{\"o}\-din\-ger's approach was electromagnetic potential components.

If we have the component $\FourierPN{\lsol{\bDB}_{0}^{lm}}{\pm|}{}$ from solution of
equation (\ref{42952668}) or (\ref{44531721}),  then we obtain immediately all components
$\FourierPN{\lsol{\bDB}_{\pm}^{lm}}{\pm|}{}$ and $\FourierPN{\lsol{\bEH}_{p}^{lm}}{\pm|}{}$ with the help of
(\ref{46557788}) and (\ref{65903676}):
\begin{equation}
\label{39785147}
\begin{array}{rlrl}
\FourierPN{\lsol{\bDB}_{0}^{lm}}{+|}{} &= \Czero_{\!+}^{lm}\,\radfppp{+}{l}{\tomega}
\;\;,&
\FourierPN{\lsol{\bDB}_{0}^{lm}}{-|}{} &= \Czero_{\!-}^{lm}\,\radfppp{-}{l}{\tomega}
\;\;,\\[2ex]
\FourierPN{\lsol{\bDB}_{+}^{lm}}{+|}{} &=
\dfrac{\Czero_{\!-}^{lm}\,\bhim\,\bigl(\rt^2\,\radfppp{-}{l}{\tomega}\bigr)_{\!;\rt}}{\sqrt{l(l+1)}\,\rt}
\;\;,&
\FourierPN{\lsol{\bDB}_{+}^{lm}}{-|}{} &=
\dfrac{\Czero_{\!+}^{lm}\,\bhim\,\bigl(\rt^2\,\radfppp{+}{l}{\tomega}\bigr)_{\!;\rt}}{\sqrt{l(l+1)}\,\rt}
\;\;,\\[2ex]
\FourierPN{\lsol{\bDB}_{-}^{lm}}{+|}{} &=
-\dfrac{\Czero_{\!-}^{lm}\,\tomega\,\sqrt{\rt^{4}+1}\,\radfppp{-}{l}{\tomega}}{\sqrt{l(l+1)}\,\rt}
\;\;,&
\FourierPN{\lsol{\bDB}_{-}^{lm}}{-|}{} &=
-\dfrac{\Czero_{\!+}^{lm}\,\tomega\,\rt^{3}\,\radfppp{+}{l}{\tomega}}{\sqrt{l(l+1)}\,\sqrt{\rt^{4}+1}}
\;\;,\\[2ex]
\FourierPN{\lsol{\bEH}_{0}^{lm}}{+|}{} &=
\dfrac{\Czero_{\!+}^{lm}\,\rt^{6}\,\radfppp{+}{l}{\tomega}}{(\rt^{4}+1)^{3/2}}
\;\;,&
\FourierPN{\lsol{\bEH}_{0}^{lm}}{-|}{} &=
\dfrac{\Czero_{\!-}^{lm}\,\rt^{2}\,\radfppp{-}{l}{\tomega}}{\sqrt{\rt^{4}+1}}
\;\;,\\[2ex]
\FourierPN{\lsol{\bEH}_{+}^{lm}}{+|}{} &=
\dfrac{\Czero_{\!-}^{lm}\,\bhim\,\sqrt{\rt^{4}+1}\,\bigl(\rt^2\,\radfppp{-}{l}{\tomega}\bigr)_{\!;\rt}}{\sqrt{l(l+1)}\,\rt^3}
\;\;,&
\FourierPN{\lsol{\bEH}_{+}^{lm}}{-|}{} &=
\dfrac{\Czero_{\!+}^{lm}\,\bhim\,\rt\,\bigl(\rt^2\,\radfppp{+}{l}{\tomega}\bigr)_{\!;\rt}}{\sqrt{l(l+1)}\,\sqrt{\rt^{4}+1}}
\;\;,\\[2ex]
\FourierPN{\lsol{\bEH}_{-}^{lm}}{+|}{} &=
-\dfrac{\Czero_{\!-}^{lm}\,\tomega\,\rt\,\radfppp{-}{l}{\tomega}}{\sqrt{l(l+1)}}
\;\;,&
\FourierPN{\lsol{\bEH}_{-}^{lm}}{-|}{} &=
-\dfrac{\Czero_{\!+}^{lm}\,\tomega\,\rt\,\radfppp{+}{l}{\tomega}}{\sqrt{l(l+1)}}
\;\;,
\end{array}
\end{equation}
where $\Czero_{\pm}^{lm}$ are complex constants, $\radfppp{\pm}{l}{\tomega}=\radfppp{\pm}{l}{\tomega}(\rt)$.

\subsection{Solutions in asymptotic power series form}
\label{apsform}
The coefficients of equations (\ref{44531721}) contain the characteristic function
\mbox{$(1+\rt^{4})^{-1}$} which can be expanded
in asymptotic power series near zero and near infinity:
\begin{equation}
\label{46345748}
\frac{1}{1+\rt^{4}} = \sum_{n=0}^{\infty}(-1)^{n} \rt^{4\, n}
= \sum_{n=1}^{\infty}(-1)^{n+1} \rt^{-4\, n}
\;\;.
\end{equation}
The convergence domain for the first series in (\ref{46345748}) is $\rt<1$, and for the second one it is
$\rt >1$. Thus we can not expect a convergence of appropriate power series
representation for the solutions for all $\rt$ values. Let us consider two asymptotic expansions.

At first consider solutions to equations (\ref{44531721}) in a generalized power series form near point of origin:
\begin{equation}
\label{49099404}
\left(\rt^2\,\radfppp{\pm}{l}{\tomega}\right) =
\rt^{\rho}\,\left(1 + \sum_{n=1}^{\infty}\Czero_n^{l,\pm}\,\rt^n\right)
\;\;,
\end{equation}
where
$\radfppp{\pm}{l}{\tomega}(\rt)$ are functions introduced in
(\ref{39785147}), $\rho$ and $\Czero_n^{l,\pm}$ ($n\geqslant 1$) must be obtained.

By direct substitution of (\ref{49099404}) into (\ref{44531721}) we have $\rho=-1$ or $\rho = 0$ for (\ref{44533615}) and
$\rho=0$ or $\rho = 3$ for (\ref{44536089}). But the cases $\rho=-1$ for (\ref{44533615}) and  $\rho=0$ for (\ref{44536089})
give rise to the singularity of types $\FourierPN{\lsol{\bDB}_{0}^{lm}}{+|}{} \sim r^{-3}$ and
$\FourierPN{\lsol{\bDB}_{-}^{lm}}{+|}{} \sim r^{-3}$, that give the appropriate divergence of an energy integral
(see also (\ref{qlllssswwe})). Thus we must take $\rho = 0$ for (\ref{44533615}) and $\rho = 3$ for (\ref{44536089}).
The coefficients $\Czero_n^{l,\pm}$ in (\ref{49099404}) can be obtained easily with the help of recurrence formulas
given by equations (\ref{44531721}) with the substitution of (\ref{49099404}).
Let us write the appropriate solutions containing several terms:
\begin{subequations}
\label{40128883}
\begin{align}
\nonumber
\radfppp{+}{l}{\tomega} &= \frac{1}{\rt^2}\left\{1 - \rt^{2}\,\frac{\tomega^2}{6}+
\rt^{4}\left[\frac{1}{20}\,l\left(l+1\right)+\frac{\tomega^4}{120}\right]
\right.
\\
\label{40132326}
&\quad
\left.- \rt^{6}\left[\frac{13\,\tomega^2}{2520}\,l\left(l+1\right) + \frac{\tomega^2}{63} +\frac{\tomega^6}{5040}\right]
+ \OOO{r^{8}}{r\to 0}\right\}
\;\;,\\
\nonumber
\radfppp{-}{l}{\tomega} &= \rt\left\{1 - \rt^{2}\,\frac{\tomega^2}{10}
+ \rt^{4}\left[\frac{1}{28}\,l\left(l+1\right) - \frac{3}{14} + \frac{\tomega^{4}}{280}\right]
\right.
\\
\label{40136314}
&\quad
\left. - \rt^{6}\left[\frac{19\,\tomega^2}{7560}\,l\left(l+1\right)
-\frac{17\,\tomega^2}{756} + \frac{\tomega^6}{15120}\right] + \OOO{r^{8}}{r\to 0}\right\}
\;\;,
\end{align}
\end{subequations}
where $\OOO{r^{n}}{r\to 0}$ denotes a quantity having the same infinitesimal order that $r^{n}$ as $r\to 0$.

Eqs. (\ref{42952668}) are converted to equation for radial spherical functions $\RadfunS{l}{\omega}$
(introduced in \cite{Chernitskii2005a})
when we take $\rb = 0$ or $r\to\infty$. These functions are connected with spherical Bessel functions.
We have the following behaviour of the radial spherical functions at infinity $r\to\infty$
(see (8.29), (8.31) in \cite{Chernitskii2005a}):
\begin{equation}
\label{36176118}
\RadfunS{l}{\omega} \sim \frac{\e{\bhim\,\omega\,r}}{\bhim^{l+1}\,r}
\;\;.
\end{equation}
Thus we can seek an appropriate asymptotic solution in the following form:
\begin{equation}
\label{42592924}
\left(\rt^2\,\radfppp{\pm}{l}{\tomega}\right) =
\bRe\left\{
\frac{\e{\bhim\,\omega\,r}}{\bhim^{l+1}}\,
\sum_{n=0}^{\infty}\Cinf_n^{l,\pm}\,\rt^{-n}
\right\}
\;\;,
\end{equation}
where $\Cinf_0^{l,\pm}\neq 0$, the complex coefficients $\Cinf_n^{l,\pm}$ for $n>0$ must be obtained,
$\bRe$ is hyperreal part. Here I take the real part
of the series to have the concordance with series (\ref{40128883})
(see (\ref{70663839}) below in subsection \ref{numsoli}).
The coefficients $\Cinf_n^{l,\pm}$ ($n>0$) can be obtained easily with the help of recurrence formulas given by
equations (\ref{44531721}) with the substitution of (\ref{42592924}). Let us write
the solution with several terms in the following form:
\begin{equation}
\label{69825122}
\radfppp{\pm}{l}{\tomega} =
\bRe\left\{
\frac{\Cinf_{\pm}^{l}}{\rt^2}\,\frac{\e{\bhim\,\omega\,r}}{\bhim^{l+1}}\left[
\sum_{l^{\prime}=0}^{l}\frac{(l+l^{\prime})!}{(l-l^{\prime})!\,l^{\prime}!}\,\left(-2\,\bhim\,\omega\,r\right)^{-l^{\prime}} \pm
\frac{1}{4\,\rt^{4}} + \dots\right]
\right\}
\;\;,
\end{equation}
where $\Cinf_{\pm}^{l}\doteqdot\Cinf_0^{l,\pm}$ of
(\ref{42592924}). Here the first term within the square brackets is characteristic for the spherical Bessel
functions.

Series (\ref{49099404}), (\ref{40128883}) must be considered as asymptotic one for sufficiently small value of $\rt$.
Series (\ref{42592924}), (\ref{69825122}) must be considered as asymptotic one for sufficiently great value of $\rt$.
Connection between these two asymptotic expansions is characterized by the complex constants
 $\Cinf_{\pm}^{l}$. These constants must be obtained with the help of another methods.
I use a numerical calculation for this purpose.

\subsection{Numerical calculation from infinity}
\label{numsoli}
Let us seek one of two solutions for each of two equations (\ref{44531721}) in the form
\begin{equation}
\label{73339643}
\frac{\e{\bhim\,\omega\,r}}{\bhim^{l+1}}\,\ypm{\pm}{l}(\rt)
\quad,
\qquad \ypm{\pm}{l}(\infty)= 1
\quad.
\end{equation}
Here the functions
$\ypm{\pm}{l}(\rt)$ have zonal index $l$, which can be omitted in the following expressions.

Change the independent variable $\rt$ to $s\doteqdot \rt^{-1}$ and introduce new unknown function
\begin{equation}
\label{58326738}
\zpm{\pm}(s) \doteqdot \ypm{\pm}{}(s^{-1})
\quad.
\end{equation}

Substitution of (\ref{73339643}) with (\ref{58326738})  into (\ref{44531721}) gives the following equations:
\begin{subequations}
\label{75527574}
\begin{align}
\label{75534838}
\zpm{+}^{''} +
  2\,\left( \frac{1}{s\,\left( 1 + s^4 \right)} -
     \frac{\bhim \,\tomega }{s^2} \right)\zpm{+}^{'}
-\frac{l\left(l + 1 \right) - 2\,\bhim \, s^3\,\tomega}{s^2\,\left( 1 + s^4 \right)}\,\zpm{+}
 &= 0
\;\;,\\
\label{75635972}
\zpm{-}^{''} +
  2\,\left( \frac{1 + 2\,s^4}{s\,\left( 1 + s^4 \right)} -
     \frac{\bhim \,\tomega }{s^2} \right)\zpm{-}^{'}
-\frac{l\left(l + 1 \right) + 2\,\bhim \,s^3\,\tomega}{s^2\,\left( 1 + s^4 \right)}\,\zpm{-}
 &= 0
\;\;.
\end{align}
\end{subequations}
These equations are investigated with the condition
$\zpm{\pm}(0)= 1$.

The numerical calculations of (\ref{75527574}) with defined value of $l$ starts from the point $s=s_1$, where
 $\zpm{\pm}(s_1)$ is calculated with the help of the power series at infinity (\ref{69825122}).
 The used value is $s_1=10^{-7}\div 10^{-8}$ and obtained results does not depend
on this value. The calculations is stopped at the point $s=s_2$.
To obtain a proper exponent of $s$ at infinity or $\rt$ at zero
for the solution, I take the value of $s_2$ to be sufficiently large: $s_2=10^{9}$.
The obtained complex numerical solutions are represented in the form
\begin{equation}
\label{42171303}
\zpm{\pm}(s) = \mmod{\zpm{\pm}}\exp{[\bhim\marg{\zpm{\pm}}]}
\quad,\qquad
\ypm{\pm}{}(\rt) = \mmod{\ypm{\pm}{}}\exp{[\bhim\marg{\ypm{\pm}{}}]}
\quad,
\end{equation}
where $\mmod{\zpm{\pm}}$ is absolute value and $\marg{\zpm{\pm}}$ is argument of $\zpm{\pm}$.

The numerical calculation gives $\mmod{\zpm{+}}\sim s$ and $\mmod{\zpm{-}}\sim 1$ as $s\to \infty$.
Using also the representation of $\ypm{\pm}{}$ as power series in $\rt$, obtained from equations (\ref{44531721}),
we have
\begin{equation}
\label{52357808}
\ypm{+}{l} = \frac{\ypma{+}{l}}{\rt}\,\e{\bhim\,\ypmp{+}{l}} + \dots
\quad,\qquad
\ypm{-}{l} = \ypma{-}{l}\,\e{\bhim\,\ypmp{-}{l}}\left(1-\bhim\,\omega\,r\right) + \dots
\quad,
\end{equation}
where $\ypma{+}{l}\doteqdot \lim\limits_{\rt\to 0}\left[\rt\,\mmod{\ypm{+}{l}(\rt)}\right]$,
$\ypma{-}{l}\doteqdot \lim\limits_{\rt\to 0}\left[\mmod{\ypm{-}{l}(\rt)}\right]$,
$\ypmp{\pm}{l}\doteqdot \lim\limits_{\rt\to 0}\left[\marg{\ypm{\pm}{l}(\rt)}\right]$.

Second linearly independent solution for each equation (\ref{44531721}) is obtained by complex conjugation
of first one (\ref{73339643}). Thus we have the following general solution to equations (\ref{44531721}):
\begin{equation}
\label{65257148}
\left(\rt^2\,\radfppp{\pm}{l}{\tomega}\right) =
\frac{1}{2}\left[\Cinf_{\pm}^{l}\,\frac{\e{\bhim\,\omega\,r}}{\bhim^{l+1}}\,\ypm{\pm}{l}(\rt)
+ {}^{\prime}\!\Cinf_{\pm}^{l}\,\frac{\e{-\bhim\,\omega\,r}}{(-\bhim)^{l+1}}\,\hconj{\ypm{\pm}{l}(\rt)}
\right]
\quad,
\end{equation}
where $\Cinf_{\pm}^{l}$ and ${}^{\prime}\!\Cinf_{\pm}^{l}$ are arbitrary complex constants.

We can obtain directly the following. In general case expression (\ref{65257148}) with (\ref{52357808}) is appropriate to
the forbidden exponents in (\ref{49099404}):
$\rho = -1$ for the ``plus'' component and $\rho = 0$ for the ``minus'' one.
To have the admissible exponents $\rho = 0$ and $\rho = 3$ accordingly, we must put
\begin{equation}
\label{45440729}
{}^{\prime}\!\Cinf_{\pm}^{l} = (-1)^{l}\,\Cinf_{\pm}^{l}\,\e{2\,\him\,\ypmp{\pm}{l}}
\;\;.
\end{equation}
Comparing also (\ref{65257148}), (\ref{45440729}) with (\ref{40128883}) I obtain
\begin{subequations}
\label{47117020}
\begin{align}
\label{47141358}
 &\Cinf_{+}^{l} = \frac{\bhim^{l}\,\e{-\bhim\,\ypmp{+}{l}}}{\ypma{+}{l}\,\tomega}
\;\;,\quad
\Cinf_{-}^{l} = \frac{3\,\bhim^{l}\,\e{-\bhim\,\ypmp{-}{l}}}{\ypma{-}{l}\,\tomega^3}
\;\;,\\
\label{47199223}
 &\mmod{\Cinf_{+}^{l}} = \frac{1}{\ypma{+}{l}\,\tomega}
\;\;,\quad
\mmod{\Cinf_{-}^{l}} = \frac{3}{\ypma{-}{l}\,\tomega^3}
\;\;,\quad
\marg{\Cinf_{\pm}^{l}} = \frac{l\,\pi}{2} - \ypmp{\pm}{l}
\;\;.
\end{align}
\end{subequations}

In view of (\ref{47141358}) relation (\ref{45440729}) gives ${}^{\prime}\!\Cinf_{\pm}^{l} = \hconj{\Cinf_{\pm}^{l}}$.
Thus we can write the following, instead of (\ref{65257148}):
\begin{equation}
\label{70663839}
\left(\rt^2\,\radfppp{\pm}{l}{\tomega}\right) =
\bRe\left\{\Cinf_{\pm}^{l}\,\frac{\e{\bhim\,\omega\,r}}{\bhim^{l+1}}\,\ypm{\pm}{l}\right\}
\;\;.
\end{equation}
It should be particularly emphasized that here we have the necessity of reality for the functions
$\radfppp{\pm}{l}{\tomega}$.
Because this I take the real parts in above-stated expressions (\ref{42592924}), (\ref{69825122}).

At infinity ($r\to\infty$) Eqs. (\ref{42952668}) take the form of equation for spherical Bessel functions.
Let us represent the solution at infinity with the help of radial spherical functions $\RadfunS{l}{\tomega}$.
Comparing (\ref{70663839}) with the appropriate definitions for $\RadfunS{l}{\tomega}$ \cite{Chernitskii2005a}
I obtain the following expression for $r\to\infty$:
\begin{equation}
\label{70679938}
\radfppp{\pm}{l}{\tomega} =
\frac{\mmod{\Cinf_{\pm}^{l}}}{\tomega^{l}\,\rt}
\left[\cos\marg{\Cinf_{\pm}^{l}}
\;\bRe\RadfunS{l}{\tomega} - \sin\marg{\Cinf_{\pm}^{l}} \;\bIm\RadfunS{l}{\tomega} \right]
\;\;,
\end{equation}
where $\mmod{\Cinf_{\pm}^{l}}$ and $\marg{\Cinf_{\pm}^{l}}$ are defined in (\ref{47199223}), $\RadfunS{l}{\tomega}\doteqdot \tomega^{l+1}\,\RadfunS{l}{}(\omega\,r)$, $\RadfunS{l}{}(\omega\,r)$ are spherical Bessel functions,
$\bRe\RadfunS{l}{}$ is the spherical Bessel function of the first kind, $\bIm\RadfunS{l}{}$
is the spherical Bessel function of the second kind.
Here, of course, $\tomega^{l}$ denotes the $l$-th power of $\tomega$.

\begin{figure}[H]
{\unitlength 1mm
\begin{picture}(110,155)
\put(0,0){
 \begin{picture}(0,0)
 \put(0,110){\includegraphics[scale=0.95]{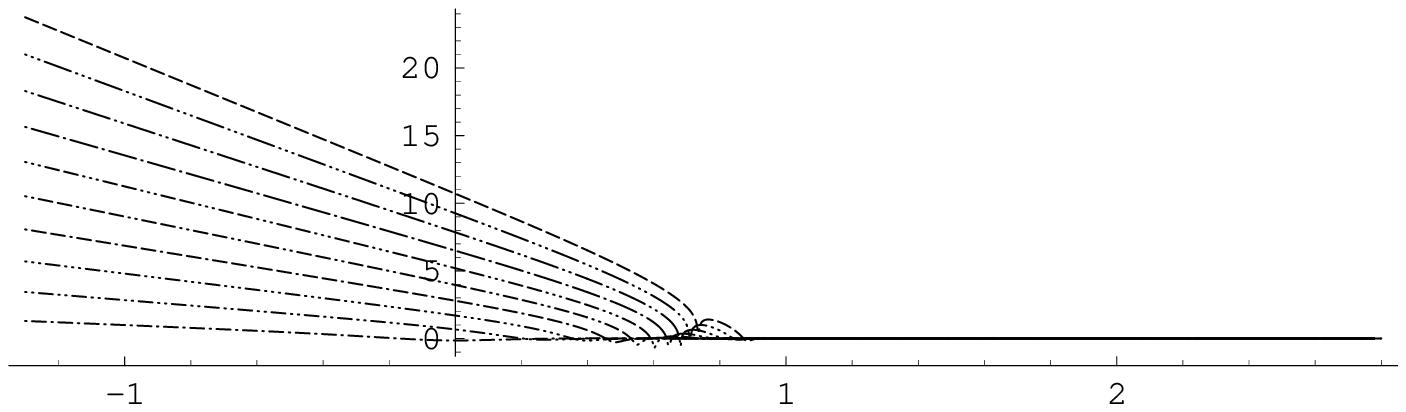}}
 \put(0,75){\includegraphics[scale=0.95]{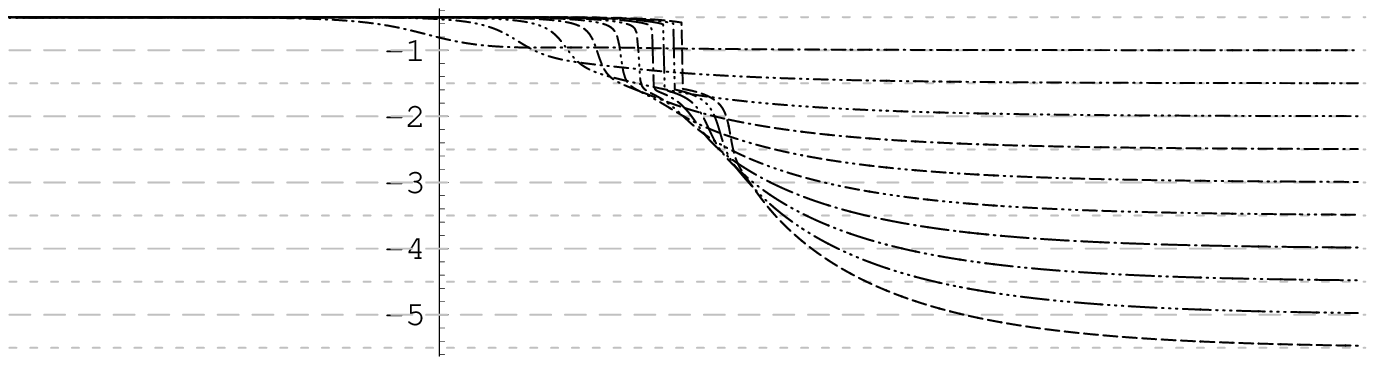}}
 \put(0,35){\includegraphics[scale=0.95]{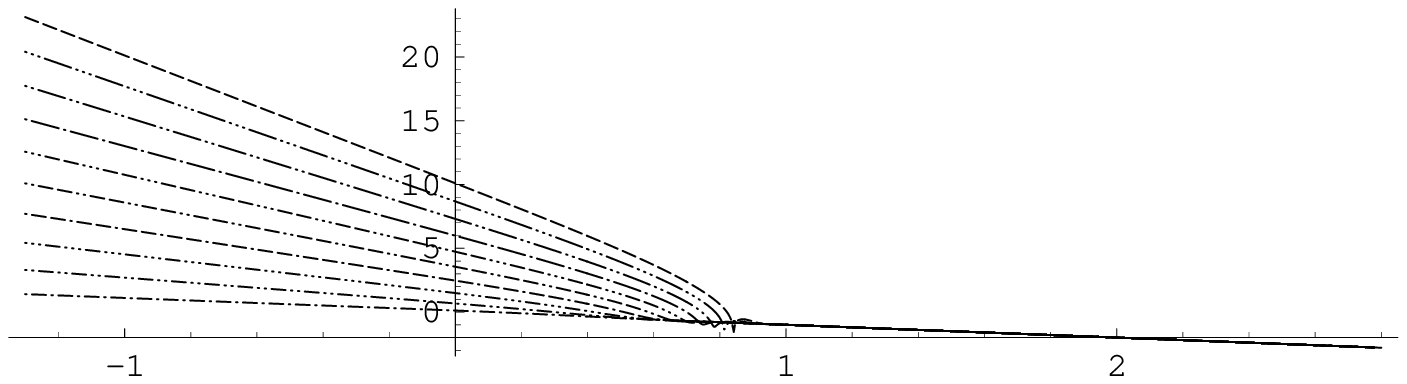}}
 \put(0,0){\includegraphics[scale=0.95]{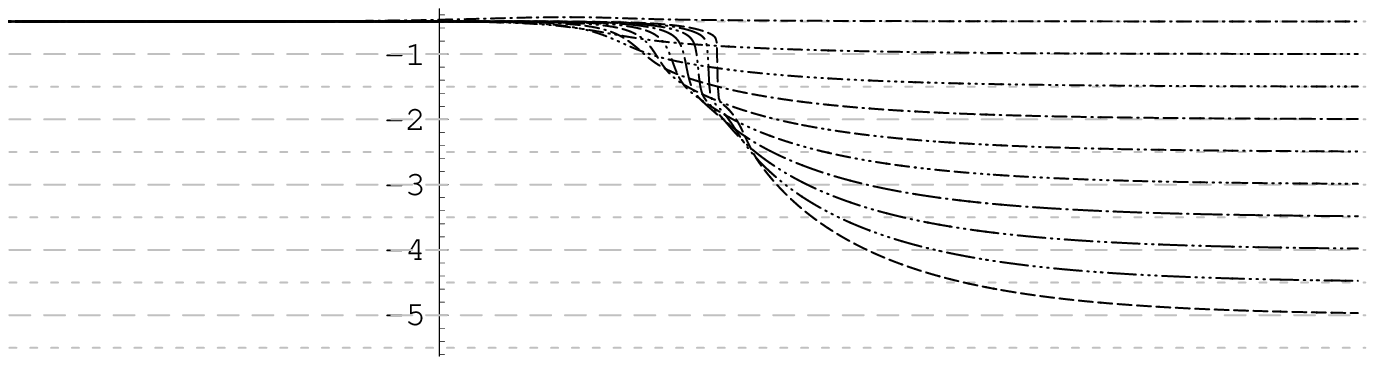}}
\put(0,119.5){\makebox(0,0)[cc]{{${\scriptstyle 1}$}}}
\put(0,122.5){\makebox(0,0)[cc]{{${\scriptstyle 2}$}}}
\put(0,125.5){\makebox(0,0)[cc]{{${\scriptstyle 3}$}}}
\put(0,128.5){\makebox(0,0)[cc]{{${\scriptstyle 4}$}}}
\put(0,131.8){\makebox(0,0)[cc]{{${\scriptstyle 5}$}}}
\put(0,135){\makebox(0,0)[cc]{{${\scriptstyle 6}$}}}
\put(0,138.5){\makebox(0,0)[cc]{{${\scriptstyle 7}$}}}
\put(0,142){\makebox(0,0)[cc]{{${\scriptstyle 8}$}}}
\put(0,145.5){\makebox(0,0)[cc]{{${\scriptstyle 9}$}}}
\put(0,149){\makebox(0,0)[cc]{{${\scriptstyle 10}$}}}
\put(0,44.5){\makebox(0,0)[cc]{{${\scriptstyle 1}$}}}
\put(0,47){\makebox(0,0)[cc]{{${\scriptstyle 2}$}}}
\put(0,49.5){\makebox(0,0)[cc]{{${\scriptstyle 3}$}}}
\put(0,52.5){\makebox(0,0)[cc]{{${\scriptstyle 4}$}}}
\put(0,55.3){\makebox(0,0)[cc]{{${\scriptstyle 5}$}}}
\put(0,58.5){\makebox(0,0)[cc]{{${\scriptstyle 6}$}}}
\put(0,61.5){\makebox(0,0)[cc]{{${\scriptstyle 7}$}}}
\put(0,65){\makebox(0,0)[cc]{{${\scriptstyle 8}$}}}
\put(0,68){\makebox(0,0)[cc]{{${\scriptstyle 9}$}}}
\put(0,71){\makebox(0,0)[cc]{{${\scriptstyle 10}$}}}
\put(134,106.5){\makebox(0,0)[cc]{{${\scriptstyle 1}$}}}
\put(134,103.5){\makebox(0,0)[cc]{{${\scriptstyle 2}$}}}
\put(134,100.5){\makebox(0,0)[cc]{{${\scriptstyle 3}$}}}
\put(134,97.2){\makebox(0,0)[cc]{{${\scriptstyle 4}$}}}
\put(134,94){\makebox(0,0)[cc]{{${\scriptstyle 5}$}}}
\put(134,90.8){\makebox(0,0)[cc]{{${\scriptstyle 6}$}}}
\put(134,87.6){\makebox(0,0)[cc]{{${\scriptstyle 7}$}}}
\put(134,84.5){\makebox(0,0)[cc]{{${\scriptstyle 8}$}}}
\put(134,81.2){\makebox(0,0)[cc]{{${\scriptstyle 9}$}}}
\put(134,78){\makebox(0,0)[cc]{{${\scriptstyle 10}$}}}
\put(134,34.6){\makebox(0,0)[cc]{{${\scriptstyle 1}$}}}
\put(134,31.4){\makebox(0,0)[cc]{{${\scriptstyle 2}$}}}
\put(134,28.3){\makebox(0,0)[cc]{{${\scriptstyle 3}$}}}
\put(134,25){\makebox(0,0)[cc]{{${\scriptstyle 4}$}}}
\put(134,21.8){\makebox(0,0)[cc]{{${\scriptstyle 5}$}}}
\put(134,18.8){\makebox(0,0)[cc]{{${\scriptstyle 6}$}}}
\put(134,15.5){\makebox(0,0)[cc]{{${\scriptstyle 7}$}}}
\put(134,12.5){\makebox(0,0)[cc]{{${\scriptstyle 8}$}}}
\put(134,9.1){\makebox(0,0)[cc]{{${\scriptstyle 9}$}}}
\put(134,5.8){\makebox(0,0)[cc]{{${\scriptstyle 10}$}}}
\put(20,149){\makebox(0,0)[cc]{{$l=1,...,10$}}}
\put(45,149){\makebox(0,0)[cl]{{$\log_{10}\ypma{+}{l}$}}}
\put(121,112.5){\makebox(0,0)[cl]{{$\log_{10}\tomega$}}}
\put(45,83){\makebox(0,0)[cl]{{$\dfrac{\ypmp{+}{l}}{\pi}$}}}
\put(45,70){\makebox(0,0)[cl]{{$\log_{10}\ypma{-}{l}$}}}
\put(121,43.5){\makebox(0,0)[cl]{{$\log_{10}\tomega$}}}
\put(45,9){\makebox(0,0)[cl]{{$\dfrac{\ypmp{-}{l}}{\pi}$}}}
 \end{picture}
 }
\end{picture}
}
\caption{The frequency dependence of amplitudes $\ypma{\pm}{l}$ and phases $\ypmp{\pm}{l}$
for $\ypm{\pm}{l}(0)$ (\ref{52357808}).}
\label{figure1}
\end{figure}

The amplitudes $\ypma{\pm}{l}$ and phases $\ypmp{\pm}{l}$ for $\ypm{\pm}{l}(0)$ (\ref{52357808}) obtained numerically for
the values $\tomega=0.05\div 600$ and $l=1\div 10$ are shown in Fig. \ref{figure1}.
These calculations give also
\begin{subequations}
\label{69655274}
\begin{align}
\label{69664108}
&
\left.
\begin{array}{ll}
\ypma{\pm}{l}= \dfrac{\ypmat{\pm}{l}}{\tomega^{l}}
\quad,\quad
&\ypmp{\pm}{l} = -\dfrac{\pi}{2}
\end{array}
\right.
\qquad\quad\quad\text{for}\quad \omega\to 0
\quad,
\\[1ex]
&
\left.
\begin{array}{ll}
\ypma{+}{l} = 1
\quad,\quad
&\ypmp{+}{l} = -\dfrac{\left(l+1\right)\pi}{2}
\\
\ypma{-}{l} = \dfrac{1}{\tomega}
\quad,\quad
&\ypmp{-}{l} = -\dfrac{l\,\pi}{2}
\end{array}
\right\}
\qquad\text{for}\quad \omega\to \infty
\quad,
\end{align}
\end{subequations}
where the coefficients $\ypmat{\pm}{l}$ in (\ref{69664108}) are rapidly increasing with increase of the zonal index $l$.
For example $\ypmat{+}{1}\approx 0.7$, $\ypmat{-}{1}\approx 1.3$,
$\ypmat{+}{5}\approx 9.1\cdot 10^{3}$, $\ypmat{-}{5}\approx 3.6\cdot 10^{3}$,
$\ypmat{+}{10}\approx 5.1\cdot 10^{10}$, $\ypmat{-}{10}\approx 1.3\cdot 10^{10}$.

\section{Linear waves around SDS}
\label{scatpw}
Any solution $\lsol{\bDB}$ can be represented in the form
$\lsol{\bDB} = \lsol{\bDB}_{-1} + \lsol{\bDB}_{0} + \lsol{\bDB}_{1}$, where each component $\lsol{\bDB}_{j}$
includes the appropriate radial functions $\FourierPN{\lsol{\bDB}_{j}^{lm}}{\pm}{}$ of expansion (\ref{46364520}).
According to (\ref{39785147}) with (\ref{37346145}) the components $\lsol{\bDB}_{\pm 1}$ are uniquely determined
by the component $\lsol{\bDB}_{0}$. Thus we can analyze this component only.
Let us write its form at infinity according to (\ref{46364520}), (\ref{37346145}), (\ref{39785147}), and (\ref{70679938}):
\begin{equation}
\label{50368979}
\lsol{\bDB}_0 = \Cn\,\sum_{l=0}^{\infty}\,\sum_{m=-l}^{l}
\spheraf{0}{l}{m}
\left[\FourierPN{\lsol{\bDB}_{0}^{lm}}{+}{}\,\e{-\bhim\,\omega\,x^0} +
\FourierPN{\lsol{\bDB}_{0}^{lm}}{-}{}\,\e{\bhim\,\omega\,x^0}\right]
\;\;,
\end{equation}
where
\vspace{-2ex}
\begin{subequations}
\label{70743453}
\begin{align}
\nonumber
\FourierPN{\lsol{\bDB}_{0}^{lm}}{+}{} &=
\frac{1}{2\,\tomega^{l}\,\rt}\left[
\Czero_{\!+}^{lm}\,\mmod{\Cinf_{+}^{l}}
\left(\cos\marg{\Cinf_{+}^{l}}
\;\bRe\RadfunS{l}{\tomega} -
\sin\marg{\Cinf_{+}^{l}}
\;\bIm\RadfunS{l}{\tomega} \right)
\right.
\\
\label{70761407}
&\quad\quad\quad\quad
+\left.
\Czero_{\!-}^{lm}\,\mmod{\Cinf_{-}^{l}}
\left(\cos\marg{\Cinf_{-}^{l}}
\;\bRe\RadfunS{l}{\tomega} -
\sin\marg{\Cinf_{-}^{l}}
\;\bIm\RadfunS{l}{\tomega} \right)
\right]
\;\;,\\
\nonumber
\FourierPN{\lsol{\bDB}_{0}^{lm}}{-}{} &=
\frac{(-1)^{m}}{2\,\tomega^{l}\,\rt}\left[
\hconj{\Czero_{\!+}^{l,-m}}\,\mmod{\Cinf_{+}^{l}}
\left(\cos\marg{\Cinf_{+}^{l}}
\;\bRe\RadfunS{l}{\tomega} -
\sin\marg{\Cinf_{+}^{l}}
\;\bIm\RadfunS{l}{\tomega} \right)
\right.
\\
\label{71687766}
&\quad\quad\quad\quad
-\left.
\hconj{\Czero_{\!-}^{l,-m}}\,\mmod{\Cinf_{-}^{l}}
\left(\cos\marg{\Cinf_{-}^{l}}
\;\bRe\RadfunS{l}{\tomega} -
\sin\marg{\Cinf_{-}^{l}}
\;\bIm\RadfunS{l}{\tomega} \right)
\right]
\;\;.
\end{align}
\end{subequations}
Here
$\Cinf_{\pm}^{l}$  are defined by (\ref{47141358}), $\Czero_{\!\pm}^{lm}$ are arbitrary complex constants.

The constants $\Czero_{\!\pm}^{lm}$ must be obtained from additional conditions.

\subsection{Linear wave scattering on SDS}
\label{response}
The known approach to the problem for wave scattering implies that an incident or acting wave is given for all space,
and a scatterer gives birth to a scattered wave.
This approach, when applied to the problem under consideration, suggests
that the field $\lsol{\bDB}$ at infinity includes two part:
\begin{equation}
\label{66568670}
\lsol{\bDB} = \lsolp{\bDB}{\,\prime} + \lsolp{\bDB}{\prime\prime}
\;\;,
\end{equation}
where $\lsolp{\bDB}{\,\prime}$ is a field of the acting (incident) wave
and $\lsolp{\bDB}{\prime\prime}$ is an additional field (scattered  or response wave), such that
the field $\lsol{\bDB}$ is a solution for problem (\ref{65903676}), (\ref{46557788}) at infinity.

The condition for distinguishing of incident and scattered waves can be their behaviour at infinity:
\begin{equation}
\label{41195755}
\mmod{\lsolp{\bDB}{\,\prime}} = \OOO{1}{r\to \infty}
\quad,\qquad
\mmod{\lsolp{\bDB}{\prime\prime}} = \OOO{r^{-1}}{r\to \infty}
\;\;.
\end{equation}
But it should be borne in mind that decomposition (\ref{66568670}) with condition (\ref{41195755})
is ambiguous.  Really, this condition remain valid if a term of higher order than $1$ ($\ooo{1}{r\to \infty}$)
is subtracted from $\lsolp{\bDB}{\prime\prime}$ and added to $\lsolp{\bDB}{\,\prime}$.

Consider the acting wave mode in the form of plane wave propagating on positive direction of
$x^{3}$ axis with arbitrary polarization. This wave can be represented as the
linear combination of two waves with mutually antithetic circular polarizations.
Using designations introduced in my article \cite{Chernitskii2005a} we have
the following representations for this wave (see Eqs. (7.28) in \cite{Chernitskii2005a}):
\begin{equation}
\label{45347999}
\lsolp{\bDB}{\,\prime} = \Cn\,\cylaf{1}{1}
\left[\polcf{+}\,\e{\bhim\,\omega\,(r\,\cos\vartheta - x^0)} +
\polcf{-}\,\e{-\bhim\,\omega\,(r\,\cos\vartheta - x^0)}\right]
\;\;,
\end{equation}
where $\omega > 0$, $\cylaf{j}{m}$ are angle cylindrical functions, relation $x^{3} = r\,\cos\vartheta$ is used,
the complex coefficient $\Cn$ is used for the concordance with (\ref{46364520}).
 Two complex numbers $\polcf{+}$ and $\polcf{-}$ (with given $\Cn$)
define a po\-la\-ri\-za\-tion ellipse and an amplitude of the wave.

According to \cite{Chernitskii2005a} (see Eqs. (8.46) in \cite{Chernitskii2005a}) we have the
following expansion
\begin{subequations}
\label{75039050}
\begin{align}
\label{43829786}
\cylaf{1}{1}\,\e{\bhim\,\omega\,r\,\cos\vartheta}
&= -\sum_{l=1}^{\infty}
\sphcoefpw{l}\,\omega^{-l-2}\garmsb{\omega}{\bumpeq}{l}{1}
\;\;,
\end{align}
where
\begin{equation}
\label{68520540}
\sphcoefpw{l} \doteqdot\bhim^{l}\,\sqrt{l\,(l+1)}\;(2\,l+1)
\end{equation}
will be called spherical coefficients of the plane wave,
\begin{align}
\label{75047633}
\garmsb{\omega}{\bumpeq}{l}{m} &\doteqdot
\FourierF{\bDB}{\bumpeq}{}_{-1}^{l}\,\spheraf{-1}{l}{m}
+\FourierF{\bDB}{\bumpeq}{}_{0}^{l}\,\spheraf{0}{l}{m}
+\FourierF{\bDB}{\bumpeq}{}_{1}^{l}\,\spheraf{1}{l}{m}
\;\;,\\
\label{75049517}
\FourierF{\bDB}{\bumpeq}{}_{0}^{l} &\doteqdot \frac{\bRe\RadfunS{l}{\omega}}{r}
\;\;,\quad
\FourierF{\bDB}{\bumpeq}{}_{\pm 1}^{l} \doteqdot \frac{1}{2}\,\sqrt{\frac{l}{l+1}}\left[
\frac{\bhim\,\omega^2}{l}\,\bRe\RadfunS{l-1}{\omega} - \left(\frac{\bhim}{r} {}\pm{} \frac{\omega}{l}\right)\bRe\RadfunS{l}{\omega}
\right]
\;\;.
\end{align}
\end{subequations}

Using representation (\ref{46364520}) for the field $\lsolp{\bDB}{\,\prime}$ (\ref{45347999})
and comparing with (\ref{75039050}), we have the following coefficients for the case
of a plane wave with elliptic polarization and propagating in positive direction of $x^{3}$ axis:
\begin{equation}
\label{41693133}
\FourierPN{\lsolp{\bDB}{\,\prime}}{\pm}{}^{lm}_{j} =
\left.-\vphantom{\frac{1}{2}}\Cron{m}{1}\,\polcf{\pm}\,
\sphcoefpw{l}\,\omega^{-l-2}\,
\FourierF{\bDB}{\bumpeq}{}_{j}^{l}\right|_{r\to\rt,\;\omega\to\pm\tomega}
\;\;,
\end{equation}
where $\Cron{m}{n}$ is the Kronecker symbol: $\Cron{m}{n}=1$ for $m=n$, $\Cron{m}{n}=0$ for $m\neq n$.
Here I use dimensionless coordinates and frequency (\ref{49802800}), $\omega\,x^{0} \equiv \tomega\,\tilde{x}^{0}$,
$\tilde{x}^{0}\doteqdot x^{0}/\rb$.

The field of plane wave expanded in terms of spherical functions (\ref{45347999}), (\ref{75039050})
includes divergent and converging elementary spherical waves \cite{Chernitskii2005a}:
\begin{equation}
\label{38313671}
\garmsb{\omega}{\pm}{l}{m}\, \e{-\bhim\, \omega\,x^0}
\;\;,
\end{equation}
where the sign index $+$ or $-$ corresponds to divergent or convergent wave accordingly.

The representation for solution as the sum of incident and scattered waves is usual for problems
of wave scattering on bodies (for electromagnetic waves see, for example, \cite{BornWolf1964,Solimeno1986}).
In this case a progressing plane wave is taken as the incident wave, and a sum of
divergent spherical waves is taken as scattered wave. Thus in this approach the scattered wave
contains the divergent spherical waves only.

According to this approach we can write:
\begin{equation}
\label{38819399}
\FourierPN{\lsol{\bDB}_{0}^{lm}}{\pm}{} = -\frac{\Cron{m}{1}\,\polcf{\pm}\,\sphcoefpw{l}\,(\pm 1)^{l}}{\tomega^{l+2}\,\rt}
\left(\bRe\RadfunS{l}{\pm\tomega} + \refC{lm}{\pm}\,\RadfunS{l}{\pm\tomega}\right)
\;\;,
\end{equation}
where the first term in the brackets corresponds to the plane wave according to (\ref{41693133}).
The second term corresponds to the divergent spherical waves (see (8.38) and below in \cite{Chernitskii2005a}),
where complex constants $\refC{lm}{\pm}$ considered as ``reflection coefficients'' of
convergent spherical waves contained in the plane wave.

Obtain radial functions $\FourierPN{\lsol{\bDB}_{0}^{l,\pm 1}}{\pm|}{}$ from (\ref{38819399}) with the help of (\ref{43850835}),
then compare it with the radial functions from (\ref{39785147}) and (\ref{70679938}). As result we have
\begin{subequations}
\label{39334667}
\begin{align}
\label{39343828}
\Czero_{\!\pm}^{l,1}\,\tomega^{2}\,\mmod{\Cinf_{\pm}^{l}}
\left[\cos\marg{\Cinf_{\pm}^{l}}
\;\bRe\RadfunS{l}{\tomega} -
\sin\marg{\Cinf_{\pm}^{l}}
\;\bIm\RadfunS{l}{\tomega} \right]
 &=
-\polcf{+}\,\sphcoefpw{l}\,
\left(\bRe\RadfunS{l}{\tomega} + \refC{l,1}{+}\,\RadfunS{l}{\tomega}\right)
\;\;,\\
\label{39346030}
\Czero_{\!\pm}^{l,-1}\,\tomega^{2}\,\mmod{\Cinf_{\pm}^{l}}
\left[\cos\marg{\Cinf_{\pm}^{l}}
\;\bRe\RadfunS{l}{\tomega} -
\sin\marg{\Cinf_{\pm}^{l}}
\;\bIm\RadfunS{l}{\tomega} \right]
 &=
\mp\hconj{\polcf{-}}\,\sphcoefpw{l}\,
\left(\bRe\RadfunS{l}{\tomega} + \hconj{\refC{l,-1}{-}}\,\RadfunS{l}{\tomega}\right)
\;\;,
\end{align}
\end{subequations}
where $\RadfunS{l}{\pm\tomega}=\bRe\RadfunS{l}{\pm\tomega} + \bhim\,\bIm\RadfunS{l}{\pm\tomega}
=\pm\bRe\RadfunS{l}{\tomega} + \bhim\,\bIm\RadfunS{l}{\tomega}$ and used $\hconj{\sphcoefpw{l}}=(-1)^{l}\,\sphcoefpw{l}$,
the constants $\Cinf_{\pm}^{l}$ are defined by (\ref{47117020}).

Because of linear independence between $\bRe\RadfunS{l}{\tomega}$ and $\bIm\RadfunS{l}{\tomega}$
we have from (\ref{39334667}) a system of linear algebraical equations for $\Czero_{\!\pm}^{l,1}$, $\refC{l,1}{+}$,
$\Czero_{\!\pm}^{l,-1}$, $\hconj{\refC{l,-1}{-}}$. Its solution is
\begin{align}
\nonumber
&\Czero_{\!\pm}^{l,1} = -\frac{\polcf{+}\,\sphcoefpw{l}}{\tomega^{2}\,\mmod{\Cinf_{\pm}^{l}}}
\,\exp[\bhim\,\marg{\Cinf_{\pm}^{l}}]
\;\;,\quad
\Czero_{\!\pm}^{l,-1} =\mp\frac{\hconj{\polcf{-}}\,\sphcoefpw{l}}{\tomega^{2}\,\mmod{\Cinf_{\pm}^{l}}}
\,\exp[\bhim\,\marg{\Cinf_{\pm}^{l}}]
\;\;,
\\[1ex]
&\refC{l,1}{+} = \hconj{\refC{l,-1}{-}} =\bhim\,\exp[\bhim\,\marg{\Cinf_{\pm}^{l}}]\,\sin\marg{\Cinf_{\pm}^{l}}
\;\;.
\label{40497631}
\end{align}
For $m\neq\pm 1$ the coefficients $\Czero_{\!\pm}^{l,m} = 0$. The solution $\FourierPN{\lsol{\bDB}_{0}^{lm}}{\pm}{}$
at infinity for linearized equation system (\ref{75771872}) is given by formula (\ref{70743453}) with coefficients
$\Czero_{\!\pm}^{l,m}$ obtained here.

The obtained solution is based on the intuitive assumption for absence of convergent waves in scattered wave mode.
But mathematically it is not quite clear why we must exclude the convergent waves from the scattered wave,
because the divergent and convergent spherical waves are complete set of functions for the representation
of solution.
However the approach considered above, when applied to the usual scattering of light on bodies, allow to obtain the
observable effect of momentum transmission from light to the scattering body.

This scattering can be called a dissipative one, because there is the permanent
energy-momentum transfer from the incident plane wave to the divergent scattered one.

\subsection{Equilibrium wave background around SDS}
\label{ewbosds}
We can also consider an experimental configuration with an equilibrium interchange between acting waves and response ones
by the scattering space region.
This situation can takes place with SDS imbedded to a wave background.
In this case the appropriate thermalization process must be accounted for by a weak wave interaction
caused by the nonlinearity of the model. This wave interaction is not considered here.

The weak wave background can be represented as expansion in terms of plane waves with various directions of propagation
and polarizations. Thus in this case we can consider also a response of SDS to one plane wave (\ref{45347999}),
 but taking into account the equilibrium between acting and response waves.
In this case the response mode $\lsolp{\bDB}{\prime\prime}$ in (\ref{66568670})
 must include divergent and convergent spherical waves, such that
there is no the permanent momentum flux through closed surface containing SDS.
This scattering can be called a non-dissipative one.

The appropriate solution at infinity for one acting plane wave, considered as a part of the weak wave background,
has the following form:
\begin{equation}
\label{48523151}
\FourierPN{\lsol{\bDB}_{0}^{lm}}{\pm}{} = -\frac{\Cron{m}{1}\,\polcf{\pm}\,\sphcoefpw{l}\,(\pm 1)^{l}}{\tomega^{l+2}\,\rt}
\left(\bRe\RadfunS{l}{\pm\tomega} + \resC{lm}{\pm}\,\bIm\RadfunS{l}{\pm\tomega}\right)
\;\;,
\end{equation}
where $\resC{lm}{\pm}$ are some coefficients.

As we see, expression (\ref{48523151}) differs from (\ref{38819399})
with change $\refC{lm}{\pm}\,\RadfunS{l}{\pm\tomega}\to\resC{lm}{\pm}\,\bIm\RadfunS{l}{\pm\tomega}$.
Thus we can consider relations (\ref{39334667}) with this change.
Here I use designations with tilde for the coefficients $\tilde{\Czero}_{\pm}^{lm}$.
Solving the appropriate system of linear algebraical equations and using
(\ref{47199223}) I obtain
\begin{subequations}
\label{39807993}
\begin{align}
\nonumber
&\tilde{\Czero}_{+}^{l1} = -\polcf{+}\,\sphcoefpw{l}\,\resfun^{l}_{+}
\;\;,\quad
\tilde{\Czero}_{-}^{l1} = -\dfrac{1}{3}\,\polcf{+}\,\sphcoefpw{l}\,\resfun^{l}_{-}
\;\;,
\\[1ex]
&\tilde{\Czero}_{+}^{l,-1} = -\hconj{\polcf{-}}\,\sphcoefpw{l}\,\resfun^{l}_{+}
\;\;,\quad
\tilde{\Czero}_{-}^{l,-1} = \dfrac{1}{3}\,\hconj{\polcf{-}}\,\sphcoefpw{l}\,\resfun^{l}_{-}
\;\;,\quad
\resC{l,1}{+} = \resC{l,-1}{-}=-\tan\marg{\Cinf_{\pm}^{l}}
\;\;,
\label{58647525}
\end{align}
where $\resfun^{l}_{\pm}$ will be called the response functions:
\begin{equation}
\label{57817523}
\resfun^{l}_{\pm} \doteqdot \frac{\tomega^{\mp 1}\,\ypma{\pm}{l}}{\cos\marg{\Cinf_{\pm}^{l}}}
\;\;.
\end{equation}
\end{subequations}

As we can see in (\ref{39807993}), the coefficients $\tilde{\Czero}_{\pm}^{l,\pm 1}$ and $\resC{l,\pm 1}{\pm}$
for solution (\ref{48523151})
have infinite values for $\cos\marg{\Cinf_{\pm}^{l}}=0$.
The condition $\cos\marg{\Cinf_{\pm}^{l}}=0$, with (\ref{47199223}) and calculations presented on Fig. \ref{figure1},
 defines resonance frequencies for the case of non-dissipative scattering.
This resonance is connected with the radius $\rb$ of SDS.

It should be noted that this resonance appears also
in the case of dissipative scattering, described in subsection \ref{response}, where $\cos\marg{\Cinf_{\pm}^{l}}=0$
is the condition for maximum of the reflection coefficients $\refC{lm}{\pm}$. In this case the coefficients of the solution
are limited because of the dissipation effect.

\begin{figure}[H]
{\unitlength 1mm
\begin{picture}(110,96)
\put(0,0){
 \begin{picture}(0,0)
 \put(-2.2,42.5){\includegraphics[scale=0.966]{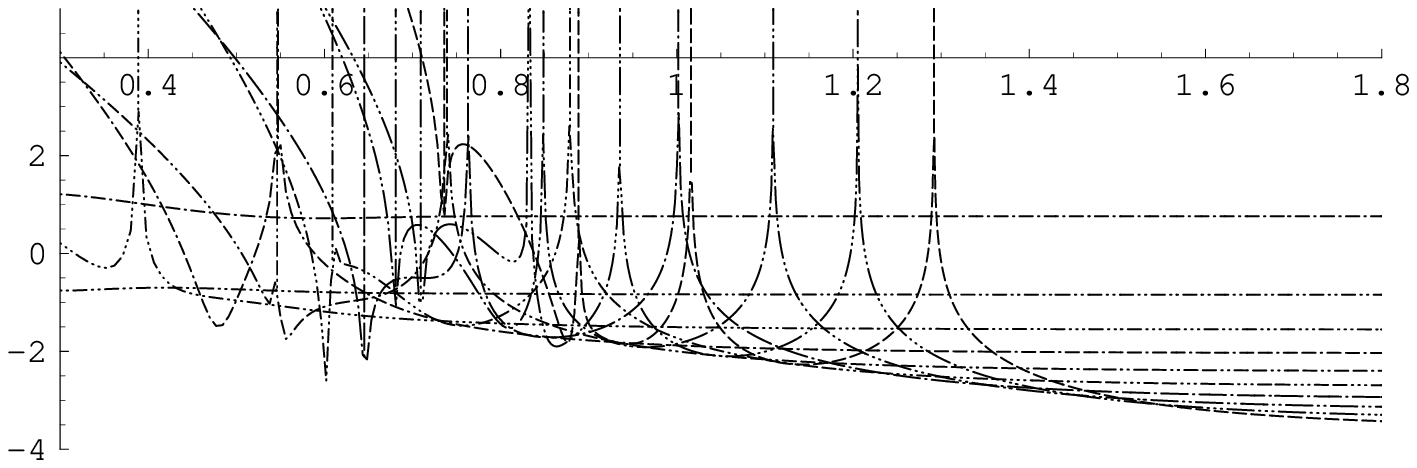}}
 \put(0,0){\includegraphics[scale=0.97]{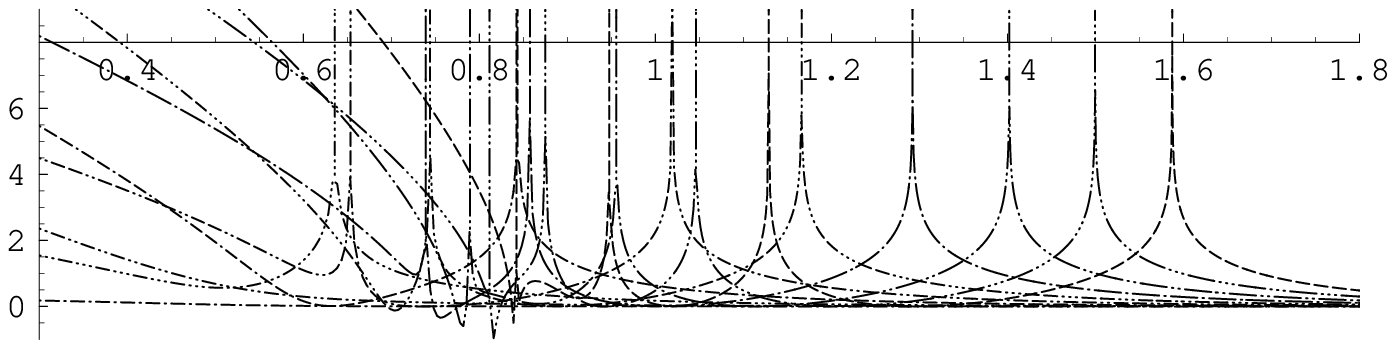}}
\put(-1,91.5){\makebox(0,0)[cl]{$\log_{10}(\resfun^{l}_{+})^2$}}
\put(1.5,79){\vector(0,1){10}}
\put(13.5,86){\makebox(0,0)[cc]{${\scriptstyle 3,}$}}
\put(25,90){\makebox(0,0)[cc]{${\scriptstyle 5,4,}$}}
\put(30.5,90){\makebox(0,0)[cc]{${\scriptstyle 6,}$}}
\put(33.5,90){\makebox(0,0)[cc]{${\scriptstyle 7,}$}}
\put(35.5,90){\makebox(0,0)[cl]{${\scriptstyle 8,9,10,5,7,}$}}
\put(49,90){\makebox(0,0)[cl]{${\scriptstyle 9,8,5,10,9,}$}}
\put(63.5,90){\makebox(0,0)[cl]{${\scriptstyle 7,10,}$}}
\put(73,90){\makebox(0,0)[cl]{${\scriptstyle 8,}$}}
\put(81,90){\makebox(0,0)[cl]{${\scriptstyle 9,}$}}
\put(90,90){\makebox(0,0)[cc]{${\scriptstyle 10}$}}
\put(125,35){\makebox(0,0)[cl]{$\log_{10}\tomega$}}
\put(-1,37.5){\makebox(0,0)[cl]{$\log_{10}(\resfun^{l}_{-})^2$}}
\put(1.5,27){\vector(0,1){8}}
\put(32,36){\makebox(0,0)[cl]{${\scriptstyle 3,5,}$}}
\put(40,36){\makebox(0,0)[cl]{${\scriptstyle 7,6,8,9,}$}}
\put(49.5,36){\makebox(0,0)[cl]{${\scriptstyle 10,4,7,9,10,8,5,9,}$}}
\put(77,36){\makebox(0,0)[cc]{${\scriptstyle 10,\,6,}$}}
\put(90,36){\makebox(0,0)[cc]{${\scriptstyle 7,}$}}
\put(98.5,36){\makebox(0,0)[cl]{${\scriptstyle 8,}$}}
\put(107,36){\makebox(0,0)[cl]{${\scriptstyle 9,}$}}
\put(114,36){\makebox(0,0)[cl]{${\scriptstyle 10}$}}
\put(125,87.5){\makebox(0,0)[cl]{$\log_{10}\tomega$}}
 \end{picture}
 }
\end{picture}
}
\caption{The frequency dependence for squares of the response functions
(\ref{57817523}). The relation between the values of zonal index $l$ and the type of dashed curve
 is the same that on Fig. \ref{figure1}. }
\label{figure5}
\end{figure}

The numerical calculation of the response functions $\resfun^{l}_{\pm}$ for $l = 1, \dots, 10$ are
shown on Fig. \ref{figure5}.
The pointed peaks on Fig. \ref{figure5} correspond to the resonance frequencies.

The values of the resonance frequencies $\tomega^{\pm}_{l n}$
are presented in Tab. \ref{44867449}, where $n$ is a number of the frequency with fixed zonal index $l$.
We can also range the resonance frequencies in ascending values. The ranged resonance frequencies up to
$\tomega^{+}_{10, 1}\approx 5.45$
are presented in Tab. \ref{44911470}. As we can see on Tab. \ref{44867449}, to obtain the correct range for
$\tomega^{\pm}_{l n} > \tomega^{+}_{10, 1}$ we must have the calculation for $l>10$.

The numerical calculations give also the following asymptotic for $\tomega\to\infty$:
\begin{equation}
\label{45270243}
\resfun^{l}_{+} \to (-1)^{l}\,\bar{\resfun}^{l}_{+}
\;\;,\quad
\resfun^{l}_{-} \to (-1)^{l}
\;\;,
\end{equation}
where $\bar{\resfun}^{l}_{+}$ can be represented by the approximate formula
\begin{equation}
\label{47536424}
\bar{\resfun}^{l}_{+} \approx \frac{1.86}{l^2} - \frac{1.91}{l^3} + \frac{2.45}{l^4}
\;\;.
\end{equation}
But if we exclude the value for $l=1$ then the good approximation is
\begin{equation}
\label{48179555}
\bar{\resfun}^{l}_{+} \approx \frac{1.52}{l^2}
\;\;.
\end{equation}
For $l=1$ this formula give a less value then one obtained with the numerical calculation
$\bar{\resfun}^{1}_{+}\approx 2.40$.

\begin{table}[H]
\begin{center}
\begin{tabular}{|p{5mm}|p{42mm}|p{42mm}|}
\hline $l$  & $\tomega^{+}_{l n}$ & $\tomega^{-}_{l n}$\\ \hline\hline
 1 & -- & --
\\
\hline
2 & -- & --
\\
\hline
3 & 2.45 & 4.32
\\
\hline
4 & {3.53}      &       {6.97}
\\
\hline
5 & {3.52, 5.48} & {4.51, 10.45}
\\
\hline
6 & {4.07, 7.56} & {5.55, 14.66}
\\
\hline
7 & {4.42, 5.79, 10.04} & {5.48, 7.20, 19.59}
\\
\hline
8 & {4.80, 7.06, 12.87} & {6.16, 9.03, 25.23}
\\
\hline
9 & {5.12, 6.80, 8.62, 16.05} & {6.48, 7.50, 11.12, 31.59}
\\
\hline
10 & 5.45, 7.73, 10.38, 19.58 & 6.95, 8.86, 13.45, 38.65
\\
\hline
\end{tabular}
\end{center}
\caption{Resonance frequencies}
\label{44867449}
\end{table}

\begin{table}[H]
\begin{center}
\begin{tabular}{|p{4mm}|p{8mm}|p{4mm}|p{4mm}|p{4mm}|p{13mm}|}
\hline $q$ & $\tomega_{q}$ &  $l$ & $n$ & $\pm $ & $\tomega_{q}/\tomega_{q-1}$\\ \hline\hline
1 & 2.45 & 3 & 1 & $+$ &
\\
\hline
2 & 3.52& 5& 1& $+$ & 1.44
\\
\hline
3 & 3.53& 4& 1& $+$ & 1.00
\\
\hline
4 & 4.07& 6& 1& $+$ & 1.15
\\
\hline
5 & 4.32& 3& 1& $-$ & 1.06
\\
\hline
6 & 4.42& 7& 1& $+$ & 1.02
\\
\hline
7 & 4.51& 5& 1& $-$ & 1.02
\\
\hline
8 & 4.80& 8& 1& $+$ & 1.07
\\
\hline
9 & 5.12& 9& 1& $+$ & 1.07
\\
\hline
10 & 5.45& 10& 1& $+$ & 1.06
\\
\hline
\end{tabular}
\end{center}
\caption{Ranged resonance frequencies}
\label{44911470}
\end{table}

We can consider a special case of the wave background with zero momentum density. In this case
the wave background can be represented as the sum of standing plane waves with
various directions of propagation and polarizations.

In this connection consider the linear response of SDS to a standing plane wave on $x^3$ axis
with arbitrary polarization.
This wave has the following form:
\begin{multline}
\label{57040322}
\lsolp{\bDB}{\,\prime} = \Cn\,\left[\polcf{+}\left(\cylaf{1}{1}\,\e{\bhim\,\omega\,x^{3}} +
\cylaf{-1}{-1}\,\e{-\bhim\,\omega\,x^{3}}\right)\e{-\bhim\,\omega\,x^0}
\right.
\\
+ \left.\polcf{-}\left(\cylaf{1}{1}\,\e{-\bhim\,\omega\,x^{3}} +
\cylaf{-1}{-1}\,\e{\bhim\,\omega\,x^{3}}\right)\e{\bhim\,\omega\,x^0}
\right]
\;\;,
\end{multline}
where $\omega > 0$, the coefficients $\polcf{+}$ and $\polcf{-}$ (with given $\Cn$)
define a po\-la\-ri\-za\-tion ellipse and an amplitude of the standing wave.

Taking the conjugation for (\ref{43829786}) and taking into account
the relations $\hconj{\sphcoefpw{l}}=(-1)^{l}\,\sphcoefpw{l}$ and
$\hconj{\!\!\garmsb{\omega}{\bumpeq}{l}{m}} = (-1)^{m}\,\garmsb{\omega}{\bumpeq}{l,}{-m}$,
I obtain
\begin{equation}
\label{53113377}
\cylaf{-1}{-1}\,\e{-\bhim\,\omega\,r\,\cos\vartheta}
= -\sum_{l=1}^{\infty}
(-1)^{l+1}\,\sphcoefpw{l}\,\omega^{-l-2}\,\garmsb{\omega}{\bumpeq}{l,}{-1}
\;\;.
\end{equation}

Thus for the case of standing wave (\ref{57040322}) we have the following coefficients of representation
(\ref{46364520}) instead of expressions (\ref{41693133}):
\begin{equation}
\label{61260620}
\FourierPN{\lsolp{\bDB}{\,\prime}}{\pm}{}^{lm}_{j} =
\left.-\left[\Cron{m}{1} + \Cron{m}{-1}\,(-1)^{l+1}\right]\polcf{\pm}\,
\sphcoefpw{l}\,\omega^{-l-2}\,
\FourierF{\bDB}{\bumpeq}{}_{j}^{l}\right|_{r\to\rt,\;\omega\to\pm\tomega}
\;\;.
\end{equation}

The coefficients of the solution for this case are
\begin{align}
\nonumber
\tilde{\Czero}_{+}^{l1} = -\left[\polcf{+} - (-1)^{l}\,\hconj{\polcf{-}}\right]\sphcoefpw{l}\,\resfun^{l}_{+}
\;\;,\quad
&\tilde{\Czero}_{-}^{l1} = -\dfrac{1}{3}\left[\polcf{+} + (-1)^{l}\,\hconj{\polcf{-}}\right]\sphcoefpw{l}\,\resfun^{l}_{-}
\;\;,
\\[1ex]
\tilde{\Czero}_{+}^{l,-1} = \left[(-1)^{l}\,\polcf{+} - \hconj{\polcf{-}}\right]\sphcoefpw{l}\,\resfun^{l}_{+}
\;\;,\quad
&\tilde{\Czero}_{-}^{l,-1} = \dfrac{1}{3}\left[(-1)^{l}\,\polcf{+} + \hconj{\polcf{-}}\right]\sphcoefpw{l}\,\resfun^{l}_{-}
\;\;.
\label{64856702}
\end{align}

\subsection{Resonance wave modes}
\label{prmodes}
The effect of wave resonance connected with two-point boundary problem is well known.
In this way the boundary condition on two resonator walls (at two points of appropriate normal coordinate)
give rise to discrete frequency spectrum
for waves within resonator, that is between the walls.

But some two-point boundary conditions can be noticed also in the problem of the scattering on SDS.
In this case we have the boundary conditions at the singular point and at infinity.
The convergence of energy integral at the origin of coordinates is the first boundary condition.
The condition at infinity is that
the resonance mode has a form which is not representable by regular harmonic waves of linear electrodynamics.

The last sentence implies that the resonance wave modes at infinity include
only imaginary part of the radial spherical functions $\bIm\RadfunS{l}{\tomega}$,
but regular spherical waves in linear electrodynamics include the appropriate real part $\bRe\RadfunS{l}{\tomega}$
only.

At the resonance frequency $\tomega=\tomega^{\pm}_{ln}$
we have $\cos\marg{\Cinf_{\pm}^{l}} = 1$, $\sin\marg{\Cinf_{\pm}^{l}} = 0$ in (\ref{70679938})
and the resonance wave modes have the following form:
\begin{equation}
\label{71591227}
{}_{\pm}\!\lsolp{\bDB}{\prime\prime}^{lnm} \doteqdot
\sum_{j=-1}^{1}\FourierPN{\lsolp{\bDB}{\prime\prime}_{j}^{ln}}{\pm}{}\,\spheraf{j}{l}{m}\,
\e{\mp\bhim\,\omega^{\pm}_{ln}\,x^0}
\;,\;\;\text{where}\;\;\;
\FourierPN{\lsolp{\bDB}{\prime\prime}_{0}^{ln}}{\pm}{} \to
\frac{\mmod{\Cinf_{\pm}^{l}}}{(\tomega^{\pm}_{ln})^{l}\,\rt}\, \bIm\RadfunS{l}{\tomega^{\pm}_{ln}}
\;\;\;\text{as}\;\;\; \rt\to\infty
\;.
\end{equation}

The resonance wave modes at infinity are orthogonal to the regular acting wave $\lsolp{\bDB}{\,\prime}$ in the following sense:
\begin{equation}
\label{64049490}
\lim\limits_{\rt_1\to\infty}\!\!\!\!\int\limits_{\rt_1}^{\rt_1 + 2\pi/\tomega}
\!\!\!\!\!\!\! \rt^2\, \df \rt
\int\limits_{0}^{\pi}\!\! \df\vartheta
\int\limits_{0}^{2\pi}\!\! \df\varphi\;
\left({}_{\pm}\!\lsolp{\bDB}{\prime\prime}^{lnm}\bcdot \hconj{\lsolp{\bDB}{\,\prime}}\right) = 0
\;\;,
\end{equation}
This orthogonality is connected with the evident relation
\begin{equation}
\label{61431716}
\lim\limits_{\rt_1\to\infty}\!\!\!\!\int\limits_{\rt_1}^{\rt_1 + 2\pi/\tomega} \!\!\!\!\!
\bRe\RadfunS{l}{\tomega}\,\bIm\RadfunS{l}{\tomega}\,\rt^2\,\df \rt
= \!\!\!
\int\limits_{\rt_1}^{\rt_1 + 2\pi/\tomega}
\!\!\!\!\!\bRe\!\left[\bhim^{-l-1}\,\e{\bhim\,\tomega\,\rt}\right]
\bIm\!\left[\bhim^{-l-1}\,\e{\bhim\,\tomega\,\rt}\right]\,\df \rt = 0
\;\;,
\end{equation}
where expression (\ref{36176118}) is used.

By using (\ref{36176118}) the boundary condition for the resonance modes at infinity can be written
also in the form:
\begin{equation}
\label{48868656}
\FourierPN{\lsolp{\bDB}{\prime\prime}_{0}^{lnm}}{\pm}{} \sim \frac{1}{\rt}\,
\bIm [\bhim^{-l-1} \exp (\bhim\,\omega^{\pm}_{ln}\, r)]
\quad\text{as}\quad \rt\to\infty
\;\;.
\end{equation}

\section{Wave solution at infinity\\and gravitational interaction}
\label{bsai}
There is a possibility for unification of electromagnetic and gravitational interactions
in the framework of the nonlinear electrodynamics because of
an effective Riemannian space effect \cite{Chernitskii1998b,Chernitskii1999,Chernitskii2002b}.
This effect implies that the energy-momentum density tensor of electromagnetic field creates the effective
Riemannian space for propagation of weak quick-oscillating waves. This effect may explain the gravitation, if
an expression of averaged energy density for electromagnetic particle contains
$\OOO{r^{-1}}{r\to \infty}$ term (see \cite{Chernitskii2002b}, the averaging is needed
for cancel quick-oscillating terms).
But for the static solution (\ref{42909272}) we have only $\OOO{r^{-4}}{r\to \infty}$ behavior.
Here I investigate this problem for representation (\ref{66568670}) in the case of non-dissipative scattering
for equilibrium wave background, stated in subsection \ref{ewbosds}.

According to (\ref{70679938}) and (\ref{39785147}) we have the following expressions for the components
of $\lsolp{\bDB}{\prime\prime}$ at $\rt \gg 1$:
\begin{subequations}
\label{42260405}
\begin{equation}
\label{39539932}
\FourierPN{\lsolp{\bDB}{\prime\prime}_{0}^{lm}}{\pm|}{} =
\resfunpp^{lm}_{\pm}\,\frac{\bIm\RadfunS{l}{\tomega}}{\rt}
\;\;,\quad
\FourierPN{\lsolp{\bDB}{\prime\prime}_{+}^{lm}}{\pm|}{} =
\dfrac{\resfunpp^{lm}_{\mp}\,\bhim\,\bigl(\rt\,\bIm\RadfunS{l}{\tomega}\bigr)_{\!;\rt}}{\sqrt{l(l+1)}\;\rt}
\;\;,\quad
\FourierPN{\lsolp{\bDB}{\prime\prime}_{-}^{lm}}{\pm|}{} =
-\dfrac{\resfunpp^{lm}_{\mp}\,\tomega\,\bIm\RadfunS{l}{\tomega}}{\sqrt{l(l+1)}}
\;\;,
\end{equation}
where
\begin{equation}
\label{42315343}
\resfunpp^{lm}_{\pm} \doteqdot
-\Czero_{\pm}^{l m}\,\mmod{\Cinf_{\pm}^{l m}}\,\tomega^{-l}\,
\sin\marg{\Cinf_{\pm}^{l}}
\;\;.
\end{equation}
\end{subequations}

Using (\ref{37346145}), (\ref{39539932}), (\ref{36176118}), and (8.34a in \cite{Chernitskii2005a}) I obtain
\begin{align}
\nonumber
\FourierPN{\lsolp{\bDB}{\prime\prime}_{\pm 1}^{lm}}{+}{} &=
\frac{\resfunpp^{lm}_{+} + \resfunpp^{lm}_{-}}{4\,\rt\,\sqrt{l\left(l+1\right)}}
\left[\mp\tomega\,\bIm\left(\bhim^{-l-1}\,\e{\bhim\,\omega\,r}\right)
- \bhim\,\bIm\left(\bhim^{-l}\,\e{\bhim\,\omega\,r}\right)\right]
+ \ooo{r^{-1}}{r\to \infty}
\;\;,
\\
\nonumber
\FourierPN{\lsolp{\bDB}{\prime\prime}_{\pm 1}^{lm}}{-}{} &=
(-1)^{m}\frac{\hconj{\resfunpp^{l,-m}_{+}} + \hconj{\resfunpp^{l,-m}_{-}}}{4\,\rt\,\sqrt{l\left(l+1\right)}}
\left[\pm\tomega\,\bIm\left(\bhim^{-l-1}\,\e{\bhim\,\omega\,r}\right)
+ \bhim\,\bIm\left(\bhim^{-l}\,\e{\bhim\,\omega\,r}\right)\right]
\\
\label{62436314}
&\quad + \ooo{r^{-1}}{r\to \infty}
\;\;,\qquad\quad
\FourierPN{\lsolp{\bDB}{\prime\prime}_{0}^{lm}}{\pm}{} = \ooo{r^{-1}}{r\to \infty}
\;\;.
\end{align}

Here I consider the case when the value of frequency $\tomega$ is near to one of its resonance values:
$\tomega\approx\tomega_{q}$
(see Table \ref{44911470}). In this case
$\tan\marg{\Cinf_{\pm}^{l}}\to \pm\infty$
and $\resfunpp^{lm}_{\pm} \to \pm\infty$ for one value of $l=l_{q}$
and definite sign index $+$ or $-$. Thus we can consider in a comparative sense
that $\resfunpp^{lm}_{\pm} \approx 0$ for $l\neq l_q$ and the appropriate sign index $+$ or $-$.

At infinity the energy-momentum density tensor has the form which is appropriate to linear electrodynamics.
In particular, the energy density has the form (\ref{alwqpk}). Thus I consider  expression $\cEp$ for $\lsol{\bDB}$
(\ref{66568670}), where I seek the terms having $ \OOO{r^{-1}}{r\to \infty}$ behavior for an averaging in
the space-time period $2\pi/\tomega$.

The four-dimensional averaging for a function $Q(x^{\mu})$ near some point $x^{\mu}$ can be defined
\begin{equation}
\label{43614770}
\average{Q}{4}{2\pi/\tomega}{5.5} \doteqdot
\frac{1}{\left(2\pi/\tomega\right)^{4}}\int\limits_{\Vol^{\prime}} Q(x^{\prime\mu})\;\dVol^{\prime}
\;\;,
\end{equation}
where $\Vol^{\prime}$ is the four-volume, such that
$x^{\mu}-\pi/\tomega \leqslant x^{\prime\mu}\leqslant x^{\mu}+\pi/\tomega$. Here we have the dependence
$\average{Q}{4}{2\pi/\tomega}{5.5} = \average{Q}{4}{2\pi/\tomega}{5.5} (x^{\mu})$.

Substitution of (\ref{66568670}) for (\ref{alwqpk}) allows to write the following expression
for the averaging energy density of the solution at infinity (\ref{66568670}):
\begin{subequations}
\label{54015502}
\begin{equation}
\label{41627430}
\average{\cEp}{4}{2\pi/\tomega}{5.5} =
\overline{\average{\cEp}{}{}{5.5}}
+ \frac{1}{8\pi\,\rt}\,\Phi^{l_q}_{\pm} (\rt,\vartheta,\varphi) + \OOO{r^{-2}}{r\to \infty}
\;\;,
\end{equation}
where
\begin{equation}
\label{48503273}
\overline{\average{\cEp}{}{}{5.5}} \doteqdot
\frac{1}{8\pi}\,\average{\,\lsolp{\bDB}{\,\prime}\bcdot\hconj{\lsolp{\bDB}{\,\prime}}\;}{4}{2\pi/\tomega}{8.5}
\;\;,
\end{equation}
cross terms with $\lsolp{\bDB}{\prime}$ and $\lsolp{\bDB}{\prime\prime}$ give
\begin{equation}
\label{42320941}
\average{\,\lsolp{\bDB}{\prime}\bcdot \hconj{\lsolp{\bDB}{\prime\prime}}
+ \hconj{\lsolp{\bDB}{\prime}} \bcdot \lsolp{\bDB}{\prime\prime}\;}{4}{2\pi/\tomega}{8.5}
= \frac{1}{\rt}\,\Phi^{l_q}_{\pm} (\rt,\vartheta,\varphi) + \OOO{r^{-2}}{r\to \infty}
\;\;.
\end{equation}
Here $\Phi^{l_q}_{\pm} (\rt,\vartheta,\varphi) = \OOO{1}{r\to \infty}$ is a function which must be obtained for each
$l_q$ and each sign index. It is evident that $\Phi_{l_q} (\rt,\pi/2,\varphi) = 0$ because the plane wave do not vary in the flat $x^{3} = 0$
and the averaging of (\ref{36176118}) gives zero for $\OOO{r^{-1}}{r\to \infty}$ terms.
Here I obtain values $\Phi_{l_q} (\rt,0,\varphi)$ and $\Phi_{l_q} (\rt,\pi,\varphi)$ that is on $x^3$ axis.
\end{subequations}

The angle spherical bivector functions is defined as
$\spheraf{j}{l}{m}(\vartheta,\varphi) = \spherzonf{j}{l}{m} \,\sphersecf{j}{m}$
\cite{Chernitskii2005a},
where $\spherzonf{j}{l}{m} = \spherzonf{j}{l}{m} (\cos\vartheta)$ are zonal functions,
$\sphersecf{j}{m} = \sphersecf{j}{m} (\vartheta,\varphi)$ are sectorial functions.
According to \cite{Vilenkin1968} we have for the zonal functions
\begin{equation}
\label{67350911}
\spherzonf{j}{l}{m} (1) = \Cron{j}{m}
\;\;,\quad
\spherzonf{j}{l}{m} (-1) = \Cron{j}{-m}\,\bhim^{2\,l} = \Cron{j}{-m}\,(-1)^{l}
\;\;,
\end{equation}
where $l$ is considered integer. According to (8.8 in \cite{Chernitskii2005a})
$\sphersecf{j}{m} (0,\varphi) = \cylaf{j}{m}$, $\sphersecf{j}{m} (\pi,\varphi) = -\cylaf{-j}{m}$
for $j=0,\pm 1$.
As result we have
\begin{equation}
\label{68749166}
\spheraf{j}{l}{m}(0,\varphi) = \Cron{m}{j}\,\cylaf{j}{m}
\;\;,\quad
\spheraf{j}{l}{m}(\pi,\varphi) = (-1)^{l+1}\,\Cron{m}{-j}\,\cylaf{-j}{m}
\;\;,
\end{equation}
where $j=0,\pm 1$.

I consider both cases as with progressing plane wave as with standing one.
 Thus I take expression (\ref{45347999}) or (\ref{57040322}) for $\lsolp{\bDB}{\,\prime}$ accordingly.
For $\lsolp{\bDB}{\prime\prime}$ I take representation
(\ref{46364520}) with coefficients (\ref{62436314}), where a number $\resfunpp^{l_q,m}_{\pm}(\tomega_q)$
 is calculated with the help of relations (\ref{42315343}) with (\ref{47117020}) and (\ref{39807993}) or (\ref{64856702}).
I substitute these expressions for $\lsolp{\bDB}{\,\prime}$, $\lsolp{\bDB}{\prime\prime}$
into (\ref{48503273}), (\ref{42320941}) and use (\ref{68749166}), (\ref{68520540}). As result we have for the progressing wave
\begin{subequations}
\label{71574191}
\begin{align}
\label{47913399}
&\overline{\average{\cEp}{}{}{5.5}} = \left(\mmod{\polcf{+}}^2 + \mmod{\polcf{-}}^2\right)/4\pi
\;\;,
\\
\label{71264166}
&\Phi^{l_q}_{\pm} (\rt,0,\varphi) = \left(2\,l_q +1\right)
\tomega^{-l_q-1}\left(\pm\mmod{\polcf{-}}^2 - \mmod{\polcf{+}}^2\right)
\,\tan\marg{\Cinf_{\pm}^{l}}
\;\;,
\\
\label{46051325}
&\Phi^{l_q}_{\pm} (\rt,\pi,\varphi) = 0
\end{align}
\end{subequations}
and for the standing wave
\begin{subequations}
\label{46034825}
\begin{align}
\label{48034391}
&\overline{\average{\cEp}{}{}{5.5}} = \left(\mmod{\polcf{+}}^2 + \mmod{\polcf{-}}^2\right)/2\pi
\;\;,
\\
\label{46037314}
&\Phi^{l_q}_{+} (\rt,0,\varphi) = \Phi^{l_q}_{+} (\rt,\pi,\varphi) = \left(2\,l_q +1\right)
\tomega^{-l_q-1}\left(\mmod{\polcf{-}}^2 - \mmod{\polcf{+}}^2\right)
\,\tan\marg{\Cinf_{+}^{l_q}}
\;\;,
\\
\nonumber
&\Phi^{l_q}_{-} (\rt,0,\varphi) = \Phi^{l_q}_{-} (\rt,\pi,\varphi)
= -\left(2\,l_q +1\right)
\tomega^{-l_q-1}\left[\left(\mmod{\polcf{-}}^2 + \mmod{\polcf{+}}^2\right)\right.
\\
\label{47566801}
&\qquad\qquad  \left. \phantom{AAAAAAAAA} +
(-1)^{l_q}\left(\polcf{+}\,\polcf{-} + \hconj{\polcf{+}}\,\hconj{\polcf{-}}\right)\right]
\,\tan\marg{\Cinf_{-}^{l_q}}
\;\;,
\end{align}
\end{subequations}
where we must take $\tan\marg{\Cinf_{+}^{l_q}}\neq 0$ or $\tan\marg{\Cinf_{-}^{l_q}}\neq 0$.

We can see in formulas (\ref{71574191}) and (\ref{46034825}) the following.
The function $\Phi^{l_q}_{\pm} (\rt,\vartheta,\varphi)$  at $x^{3}$ axis is asymmetrical
for the case with progressing wave (\ref{71574191}) and it is
 symmetrical for the case with standing wave (\ref{46034825}).
The values of the function $\Phi^{l_q}_{+} (\rt,\vartheta,\varphi)$ at $x^3$
for the standing wave can be obtained by
summation of the appropriate values for progressing waves with opposite directions of propagation
(compare (\ref{71264166}) and (\ref{46051325}) with (\ref{46037314})). But
the values of the function $\Phi^{l_q}_{-} (\rt,\vartheta,\varphi)$ at $x^3$
 can not be obtained by the summation
(compare (\ref{71264166}) and (\ref{46051325}) with (\ref{47566801})).
The function $\Phi^{l_q}_{\pm} (\rt,\vartheta,\varphi)$ can take positive or negative values.
It should be noted that here we have the behavior which is typical for interference of waves
($\lsolp{\bDB}{\,\prime}$ and $\lsolp{\bDB}{\prime\prime}$ for this case).

The gradient of the averaged energy density creates an appropriate co-directed force for another electromagnetic
particles and distorts rays of light. This effect appear because
the effective metric component contains the averaged energy density \cite{Chernitskii1998b,Chernitskii1999,Chernitskii2002b}:
\begin{equation}
\label{67374392}
\metrEff_{00} = \metr_{00} - 4\pi\,\chi^2\,\average{\cEp}{}{}{5.5}
\;\;.
\end{equation}

As we see in (\ref{71574191}) and (\ref{46034825}), the near resonance wave solution with one acting plane wave can give both
attraction ($\Phi^{l_q}_{\pm}>0$) and repulsion ($\Phi^{l_q}_{\pm} < 0$) effects. Also there is an angular dependence.
But it would appear reasonable that the angular dependence can be canceled
with taking into consideration of all acting plane waves contained in the equilibrium wave background,
as well as by an averaging for a bulk of
electromagnetic particles just as in macroscopic bodies. The appropriate averaged energy density can have the form
(\ref{41627430}) with substitution $\Phi^{l_q}_{\pm}\to \Phi = \text{\textit{Constant}}$.
In principle, the following three cases is possible: $\Phi > 0$, $\Phi = 0$, and $\Phi < 0$. The case $\Phi > 0$
corresponds to attraction which is essential for explanation of gravitation. However, in the context of such
approach we must not exclude a possibility of special and artificial field configurations which give the
repulsion or antigravitation effect, for example, with some coherent states of matter.

\section{Wave solution at zero\\and beyond linearization}
\label{ressquant}
The radial functions $\radfppp{+}{l}{\tomega}$ and $\radfppp{-}{l}{\tomega}$ contained in solution (\ref{39785147})
have different behavior near zero.
The functions $\radfppp{+}{l}{\tomega}$  has
the main term proportional to $\rt^{-2}$ (\ref{40132326}), just like basic solution (\ref{42909272}).
But the main term of the functions $\radfppp{-}{l}{\tomega}$ is proportional to $\rt$ (\ref{40136314}).

If we have the resonance increase for the amplitudes $\Czero^{lm}_{\pm}$ of the solution
(see (\ref{39807993}) and (\ref{64856702})) then linearization (\ref{39606993}) is no
longer well approximation. But the plus component resonance $\Czero^{l}_{+}$
will rather destroy the linearization then the minus component one $\Czero^{l}_{-}$,
because of the different behaviour for the components near zero.

Thus we must regard the minus components resonance frequencies $\tomega_{q}^{-}$ to be more
plausible then the plus ones $\tomega_{q}^{+}$ (see Tab. \ref{44911470}).

In any case, to avoid the infinite amplitude $\Czero^{l}_{\pm}$ at a resonance frequency
we must consider a nonlinear resonance by taking into account
highest terms in expansion (\ref{39606993}) or solving the problem with original constitutive relation
(\ref{63244197}).

It is known, in general resonance phenomena theory for oscillating system, that resonance
frequencies for the case of nonlinear resonance most often depend on an influence amplitude.
But this dependence can be weak or strong.
What is more, we can suppose that some resonance influences will destroy the singularity under consideration
and will give birth for singularities of another types.

It should be noted also that a veritable wave solution of the suitable nonlinear problem can
essentially differ from the wave solution of the linearized problem.

\section{Conclusions}
\label{concl}
The present investigation gives the method and solutions concerning the problem linearized around
the static dyon solution for Born-Infeld nonlinear electrodynamics. In particular, the
wave resonance effect on static electromagnetic particle is discovered. This effect appears explicitly, if
the the static electromagnetic particle is imbedded to an equilibrium electromagnetic wave background.

But we must consider an electromagnetic wave background to be always in space.
Thus we can not have a static solution without the appropriate wave modes.
We can suppose that the wave modes will essentially influence on behavior and
manifestations of the particle solution in its interaction with another particles.
The demonstrated possible connection with the gravitational interaction is one of such
manifestations.

\appendix

\renewcommand{\theequation}{\Alph{section}.\arabic{equation}}

\section{Main designations}
\label{STELRESdst}
\vspace{-1.5ex}
\begin{tabular*}{\textwidth}{|p{13.5ex}|l@{\extracolsep{\fill}\hspace*{2mm}}|c|}
\hline \vspace{-7pt} Symbol & \parbox[t]{41.5ex}{\vspace{-7pt}Name} & \parbox[t]{11.5ex}{\vspace{-7pt} Appearance}\\
\hline
\hline \vspace{-7pt} $\metr_{\mu\nu}$, $\metrEff_{\mu\nu}$ \vspace{2pt} & \parbox[t]{41.5ex}{\vspace{-7pt} Metric tensor, effective metric\vspace{2pt}} & \parbox[t]{11.5ex}{\vspace{-7pt}(\ref{dkmfjysg}), (\ref{67374392})\vspace{2pt}}\\
\hline \vspace{-7pt} $\bcdot$, $\bwedge$ \vspace{2pt} & \parbox[t]{41.5ex}{\vspace{-7pt} Symmetrical and asymmetrical products\vspace{2pt}} & \parbox[t]{11.5ex}{\vspace{-7pt} (\ref{dkmfjysg})\vspace{2pt}}\\
\hline \vspace{-7pt} $\baab^\mu$, $\bbaab^i$ \vspace{2pt} & \parbox[t]{41.5ex}{\vspace{-7pt} Basis vectors and bivectors\vspace{2pt}} & \parbox[t]{11.5ex}{\vspace{-7pt} (\ref{dkmfjysg})\vspace{2pt}}\\
\hline \vspace{-7pt} $\bEem$, $\Eem_i$, $\bHem$, $\Hem_i$ \vspace{2pt} & \parbox[t]{41.5ex}{\vspace{-7pt} Intensities of  fields: electric and magnetic\vspace{2pt}} & \parbox[t]{11.5ex}{\vspace{-7pt} (\ref{YZ})\vspace{2pt}}\\
\hline \vspace{-7pt} $\bDem$, $\Dem^i$, $\bBem$, $\Bem^i$%
& \parbox[t]{41.5ex}{\vspace{-7pt} Inductions: electric and magnetic\vspace{2pt}} & \parbox[t]{11.5ex}{\vspace{-7pt} (\ref{YZ})\vspace{2pt}}\\
\hline \vspace{-7pt} $\bDB$, $\bEH$ \vspace{2pt} & \parbox[t]{41.5ex}{\vspace{-7pt} Quasi-bivectors of electromagnetic field\vspace{2pt}} & \parbox[t]{11.5ex}{\vspace{-7pt} (\ref{YZ})\vspace{2pt}}\\
\hline \vspace{-7pt} $\hconj{\mathbf{C}}$ \vspace{2pt} & \parbox[t]{41.5ex}{\vspace{-7pt} Hyperconjugation\vspace{2pt}} & \parbox[t]{11.5ex}{\vspace{-7pt} (\ref{44468525})\vspace{2pt}}\\
\hline \vspace{-7pt} $\cE$, $\cEt$, $\cEp$, $\bcP$, $\cP_i$ %
& \parbox[t]{41.5ex}{\vspace{-7pt} Energy density and momentum density\vspace{2pt}} & \parbox[t]{11.5ex}{\vspace{-7pt} (\ref{qlllssswwe})\vspace{2pt}}\\
\hline \vspace{-7pt} $\bsoll{\bDB}$, $\lsol{\bDB}$, $\lsolp{\bDB}{\prime\prime}$ \vspace{2pt} & \parbox[t]{41.5ex}{\vspace{-7pt} Basic solution for linearization and linearized solution\vspace{2pt}} & \parbox[t]{11.5ex}{\vspace{-7pt} (\ref{36671462}), (\ref{66568670})\vspace{2pt}}\\
\hline \vspace{-7pt} $\C$, $\Cn$, $\Ce$, $\Cm$ \vspace{2pt} & \parbox[t]{41.5ex}{\vspace{-7pt} Electromagnetic, normed, electric, and magnetic charges\vspace{2pt}} & \parbox[t]{11.5ex}{\vspace{-7pt} (\ref{42909272})\vspace{2pt}}\\
\hline \vspace{-7pt} $\rb$, $\rt$, $\tomega$ \vspace{2pt} & \parbox[t]{41.5ex}{\vspace{-7pt} Radius of SDS, dimensionless radius and frequency\vspace{2pt}} & \parbox[t]{11.5ex}{\vspace{-7pt} (\mbox{\ref{42909272}), (\ref{49802800})}\vspace{2pt}}\\
\hline \vspace{-7pt} $\spheraf{j}{l}{m}$, $\sphersecf{j}{m}$, $\spherzonf{j}{l}{m}$ \vspace{2pt} & \parbox[t]{41.5ex}{\vspace{-7pt} Angle spherical functions: general, sectorial, and zonal\vspace{2pt}} & \parbox[t]{11.5ex}{\vspace{-7pt} \mbox{(\ref{46364520}), (\ref{67350911})}\vspace{2pt}}\\
\hline \vspace{-7pt} $\Czero^{lm}_{\pm}$, $\Cinf^{lm}_{\pm}$ \vspace{2pt} & \parbox[t]{41.5ex}{\vspace{-7pt} Constants for the linearized solution\vspace{2pt}} & \parbox[t]{11.5ex}{\vspace{-7pt} \mbox{\!\!(\ref{39785147}), (\ref{69825122})}\vspace{2pt}}\\
\hline \vspace{-7pt} $\radfppp{\pm}{l}{\tomega}$ \vspace{2pt} & \parbox[t]{41.5ex}{\vspace{-7pt} Radial functions of the problem\vspace{2pt}} & \parbox[t]{11.5ex}{\vspace{-7pt} (\ref{39785147})\vspace{2pt}}\\
\hline \vspace{-7pt} $\OOO{r^{-1}}{r\to \infty}$, $\ooo{r^{-1}}{r\to \infty}$ \vspace{2pt} & \parbox[t]{41.5ex}{\vspace{-7pt} Infinitesimals of the same and higher order\vspace{2pt}} & \parbox[t]{11.5ex}{\vspace{-7pt} \mbox{(\ref{40128883}), (\ref{62436314})}\vspace{2pt}}\\
\hline \vspace{-7pt} $\bRe\mathbf{C}$, $\bIm\mathbf{C}$ \vspace{2pt} & \parbox[t]{41.5ex}{\vspace{-7pt} Hyperreal and hyperimaginary parts\vspace{2pt}} & \parbox[t]{11.5ex}{\vspace{-7pt} \mbox{\!\!(\ref{42592924}), (\ref{70679938})}\vspace{2pt}}\\
\hline \vspace{-7pt} $\RadfunS{l}{k_r}$ \vspace{2pt} & \parbox[t]{41.5ex}{\vspace{-7pt} Radial spherical functions:\vspace{2pt}} & \parbox[t]{11.5ex}{\vspace{-7pt} \mbox{\!\!(\ref{36176118}), (\ref{70679938})}\vspace{2pt}}\\
\hline \vspace{-7pt} $\mmod{C}$, $\marg{C}$ \vspace{2pt} & \parbox[t]{41.5ex}{\vspace{-7pt} Modulus and argument of hyperscalar\vspace{2pt}} & \parbox[t]{11.5ex}{\vspace{-7pt} (\ref{42171303})\vspace{2pt}}\\
\hline \vspace{-7pt} $\cylaf{j}{m}$ \vspace{2pt} & \parbox[t]{41.5ex}{\vspace{-7pt} Angle cylindrical functions\vspace{2pt}} & \parbox[t]{11.5ex}{\vspace{-7pt} (\ref{45347999})\vspace{2pt}}\\
\hline \vspace{-7pt} $\polcf{\pm}$ \vspace{2pt} & \parbox[t]{41.5ex}{\vspace{-7pt} Polarization coefficients\vspace{2pt}} & \parbox[t]{11.5ex}{\vspace{-7pt} \mbox{(\ref{45347999}), (\ref{57040322})}\vspace{2pt}}\\
\hline \vspace{-7pt} $\sphcoefpw{l}$ \vspace{2pt} & \parbox[t]{41.5ex}{\vspace{-7pt} Spherical coefficients of plane wave\vspace{2pt}} & \parbox[t]{11.5ex}{\vspace{-7pt} (\ref{68520540})\vspace{2pt}}\\
\hline \vspace{-7pt} $\Cron{m}{n}$ \vspace{2pt} & \parbox[t]{41.5ex}{\vspace{-7pt} Kronecker symbol\vspace{2pt}} & \parbox[t]{11.5ex}{\vspace{-7pt} (\ref{41693133})\vspace{2pt}}\\
\hline \vspace{-7pt} $\resfun^{l}_{\pm}$, $\resfunpp^{l}_{\pm}$ \vspace{2pt} & \parbox[t]{41.5ex}{\vspace{-7pt} Response functions\vspace{2pt}} & \parbox[t]{11.5ex}{\vspace{-7pt} \mbox{\!\!\!\!(\ref{57817523}), (\ref{42315343})}\vspace{2pt}}\\
\hline \vspace{-7pt} $\average{Q}{4}{2\pi/\tomega}{5.5}$ \vspace{2pt} & \parbox[t]{41.5ex}{\vspace{-7pt} Averaging operation\vspace{2pt}} & \parbox[t]{11.5ex}{\vspace{-7pt} (\ref{43614770})\vspace{2pt}}\\
\hline
\end{tabular*}


\begin{thebibliography}{10}
\addcontentsline{toc}{section}{\refname}

\providecommand{\url}[1]{\texttt{#1}}
\providecommand{\urlprefix}{URL }
\expandafter\ifx\csname urlstyle\endcsname\relax
  \providecommand{\doi}[1]{doi:\discretionary{}{}{}#1}\else
  \providecommand{\doi}{doi:\discretionary{}{}{}\begingroup
  \urlstyle{rm}\Url}\fi
\providecommand{\bibinfo}[2]{#2}
\providecommand{\eprint}[2][]{\url{#2}}

\bibitem{BornWolf1964}
\bibinfo{author}{M.~Born} and \bibinfo{author}{E.~Wolf}.
\newblock \emph{\bibinfo{title}{Principles of optics}}.
\newblock \bibinfo{publisher}{Pergamon Press}, \bibinfo{year}{1964}.

\bibitem{Chernitskii1998b}
\bibinfo{author}{A.~A. Chernitskii}.
\newblock \bibinfo{title}{Light beams distortion in nonlinear electrodynamics}.
\newblock \emph{\bibinfo{journal}{J. High Energy Phys.}}
  \textbf{\bibinfo{volume}{1998}}(\bibinfo{number}{11}) \bibinfo{pages}{Paper
  15 , 1--5}, \bibinfo{month}{November} \bibinfo{year}{1998}.
\newblock \href{http://xxx.lanl.gov/abs/hep-th/9809175}{{\tt hep-th/9809175}}.

\bibitem{Chernitskii1999}
\bibinfo{author}{A.~A. Chernitskii}.
\newblock \bibinfo{title}{Dyons and interactions in nonlinear
  (\uppercase{B}orn-\uppercase{I}nfeld) electrodynamics}.
\newblock \emph{\bibinfo{journal}{J. High Energy Phys.}}
  \textbf{\bibinfo{volume}{1999}}(\bibinfo{number}{12}) \bibinfo{pages}{Paper
  10, 1--34}, \bibinfo{year}{1999}.
\newblock \href{http://xxx.lanl.gov/abs/hep-th/9911093}{{\tt hep-th/9911093}}.

\bibitem{Chernitskii2002a}
\bibinfo{author}{A.~A. Chernitskii}.
\newblock \bibinfo{title}{Born-\uppercase{I}nfeld electrodynamics:
  \uppercase{C}lifford number and spinor representations}.
\newblock \emph{\bibinfo{journal}{Int. J. Math. \& Math. Sci.}}
  \textbf{\bibinfo{volume}{31}}(\bibinfo{number}{2}) \bibinfo{pages}{77--84},
  \bibinfo{year}{2002}.
\newblock \href{http://xxx.lanl.gov/abs/hep-th/0009121}{{\tt hep-th/0009121}}.

\bibitem{Chernitskii2002b}
\bibinfo{author}{A.~A. Chernitskii}.
\newblock \bibinfo{title}{Induced gravitation as nonlinear electrodynamics
  effect}.
\newblock \emph{\bibinfo{journal}{Gravitation \& Cosmology}}
  \textbf{\bibinfo{volume}{8}} \bibinfo{pages}{Supplement, 157--160},
  \bibinfo{year}{2002}.
\newblock \href{http://xxx.lanl.gov/abs/gr-qc/0211034}{{\tt gr-qc/0211034}}.

\bibitem{Chernitskii2004a}
\bibinfo{author}{A.~A. Chernitskii}.
\newblock \bibinfo{title}{Born-\uppercase{I}nfeld equations}.
\newblock In \emph{\bibinfo{booktitle}{Encyclopedia of \uppercase{N}onlinear
  \uppercase{S}cience}}, edited by \bibinfo{editor}{A.~Scott}, pages
  \bibinfo{pages}{67--69}. \bibinfo{publisher}{Routledge},
  \bibinfo{address}{New York and London}, \bibinfo{year}{2004}.
\newblock \href{http://xxx.lanl.gov/abs/hep-th/0509087}{{\tt hep-th/0509087}}.

\bibitem{Chernitskii2005a}
\bibinfo{author}{A.~A. Chernitskii}.
\newblock \bibinfo{title}{Basic systems of orthogonal functions for space-time
  multivectors}.
\newblock \emph{\bibinfo{journal}{Advances in applied Clifford algebras}}
  \textbf{\bibinfo{volume}{15}}(\bibinfo{number}{1}) \bibinfo{pages}{27--53},
  \bibinfo{year}{2005}.
\newblock \href{http://xxx.lanl.gov/abs/hep-th/0501161}{{\tt hep-th/0501161}}.

\bibitem{HutsonPym1980}
\bibinfo{author}{V.~C.~L. Hutson} and \bibinfo{author}{J.~S. Pym}.
\newblock \emph{\bibinfo{title}{Applications of \uppercase{F}unctional
  \uppercase{A}nalysis and \uppercase{O}perator \uppercase{T}heory}}.
\newblock \bibinfo{publisher}{Academic Press}, \bibinfo{address}{London},
  \bibinfo{year}{1980}.

\bibitem{Schrodinger1942a}
\bibinfo{author}{E.~Schr{\"o}dinger}.
\newblock \bibinfo{title}{Non-linear optics}.
\newblock \emph{\bibinfo{journal}{Proc. Roy. Irish Acad. A}}
  \textbf{\bibinfo{volume}{47}} \bibinfo{pages}{77--117}, \bibinfo{month}{June}
  \bibinfo{year}{1942}.

\bibitem{Schrodinger1942b}
\bibinfo{author}{E.~Schr{\"o}dinger}.
\newblock \bibinfo{title}{Dynamics and scattering-power of \uppercase{B}orn's
  electron}.
\newblock \emph{\bibinfo{journal}{Proc. Roy. Irish Acad. A}}
  \textbf{\bibinfo{volume}{48}} \bibinfo{pages}{91--122},
  \bibinfo{month}{November} \bibinfo{year}{1942}.

\bibitem{Solimeno1986}
\bibinfo{author}{S.~Solimeno}, \bibinfo{author}{B.~Crosignani}, and
  \bibinfo{author}{P.~DiPorto}.
\newblock \emph{\bibinfo{title}{Guiding, \uppercase{D}iffraction, and
  \uppercase{C}onfinement of \uppercase{O}ptical \uppercase{R}adiation}}.
\newblock \bibinfo{publisher}{Academic Press, Inc}, \bibinfo{year}{1986}.

\bibitem{Vilenkin1968}
\bibinfo{author}{N.~Y. Vilenkin}.
\newblock \emph{\bibinfo{title}{Special functions and the theory of group
  representations}}.
\newblock Translations of Mathematical Monographs. \bibinfo{publisher}{AMS},
  \bibinfo{year}{1968}.

\end{thebibliography}
\end{document}